\documentclass{ws-procs975x65}

\newcommand{\be}[1]{\begin{equation}\label{#1}}
\newcommand{\beq}{\begin{equation}}
\newcommand{\eeq}{\end{equation}}
\newcommand{\beqn}[1]{\begin{eqnarray}\label{#1}}
\newcommand{\eeqn}{\end{eqnarray}}

\newcommand{\dub}[2]{\left(\begin{array}{c}{#1}\\{#2}
\end{array}\right)}

\newcommand{\mat}[4]{\left(\begin{array}{cc}{#1}&{#2}\\{#3}&{#4}
\end{array}\right)}

\renewcommand{\to}{\rightarrow}
\def\tolr{\leftrightarrow}

\def\ov{\overline}
\def\ee{\end{equation}}

\def\lsim{\raise0.3ex\hbox{$\;<$\kern-0.75em\raise-1.1ex
\hbox{$\sim\;$}}}
\def\gsim{\raise0.3ex\hbox{$\;>$\kern-0.75em\raise-1.1ex
\hbox{$\sim\;$}}}

\def\bg{{\bf g}}

\def\cal{\mathcal}

\def\cG{{\cal G}}

\def\cL{{\cal L}}

\def\al{\alpha}
\def\ga{\gamma}
\def\Ga{\Gamma}

\def\la{\lambda}
\def\La{\Lambda}  

\def\Om{\Omega}
\def\eps{\varepsilon}

\def\DN{\Delta N_\nu}

\def\rhb{\bar{\rho}}

\def\bg{\bar{g}_\ast}

\def\tphi{\tilde{\phi}}

\def\tf{\tilde{f}}

\def\tq{\tilde{q}}
\def\tl{\tilde{l}}
\def\tL{\tilde{L}}
\def\tR{\tilde{R}}
\def\tN{\tilde{N}}
\def\te{\tilde{e}}
\def\tu{\tilde{u}}
\def\td{\tilde{d}}
\def\tnu{\tilde{\nu}}

\def\lpr{l^\prime}
\def\phpr{\phi^\prime}

\def\barl{\bar{l}}
\def\barphi{\bar{\phi}}
\def\barlpr{\bar{l}^\prime}
\def\barphpr{\bar{\phi}^\prime}
\def\olga{{\it Olga}}
\def\maxim{{\it Maxim}}
\def\alice{{\it Alice}}

\begin{document}


\title{
THROUGH THE LOOKING-GLASS: ALICE'S ADVENTURES \\
IN MIRROR WORLD }

\author{ZURAB BEREZHIANI}
\address{
 Dipartimento di Fisica, Universit\`a di L'Aquila, 67010 Coppito,
L'Aquila, and \\
INFN, Laboratori Nazionali del Gran Sasso, 67010 Assergi, L'Aquila,
Italy; \\
e-mail: berezhiani@fe.infn.it
}



\maketitle


\abstracts{
We briefly review the concept of a parallel `mirror' world
which has the same particle physics as the observable world
and couples to the latter by gravity and perhaps other very 
weak forces. 
The nucleosynthesis bounds demand that the mirror world
should have a smaller temperature than the ordinary one.
By this reason its evolution should substantially deviate 
from the standard cosmology as far as the crucial epochs like
baryogenesis, nucleosynthesis etc. are concerned.
In particular, we show that in the context of certain 
baryogenesis scenarios, the baryon asymmetry in the mirror 
world should be larger than in the observable one. 
Moreover, we show that mirror baryons could naturally 
constitute the dominant dark matter component of the Universe, 
and discuss its cosmological implications.  
} 

\vspace{10mm} 

Published in Ian Kogan Memorial Collection 
{\it ``From Fields to Strings: Circumnavigating Theoretical Physics"}, 
Eds. M. Shifman et al., World Scientific, Singapore, 
vol. 3,  pp. 2147-2195. 




\newpage 

{\sl `Now, if you'll only attend, Kitty, and not talk so much,
I'll tell you all my ideas about Looking-glass House. 
First, there's the room you can see through the glass -- 
that's just the same as our drawing-room, only the things go 
the other way... 
the books are something like our books, 
only the words go the wrong way: 
I know that, because I've held up one of our books to the glass, 
and then they hold up one in the other room.
I can see all of it when when I get upon a chair --
all but the bit just behind the fireplace. 
I do so wish I could see that bit! 
I want so to know whether they've a fire in the winter: 
you never can tell, you know, unless our fire smokes, 
and then smoke comes up in that room too -- 
but that may be only pretence, 
just to make it look as if they had a fire... 

`How would you like to leave in the Looking-glass House, 
Kitty? I wander if they'd give you milk in there? 
Perhaps Looking-glass milk isn't good to drink -- but Oh, Kitty!
Now we come to the passage. You can just see a little peep 
of the passage in Looking-glass House, if you leave the door of our 
drawing-room wide open: and it's very like our passage as far as 
you can see, only you know it may be quite on beyond. 
Oh, Kitty, how how nice it would be if we could get through 
into Looking-glass House! Let's pretend there's a way of getting 
through into it, somehow, Kitty... Why, it's turning into a sort 
of mist now, I declare! It'll be easy enough to get through -- 
'She said this, though she hardly knew how she had got there...

In another moment Alice was through the glass, and had jumped 
lightly down into the Looking-glass room. 
The very first thing Alice did was to look whether there was 
a fire in the fireplace, and she was quite pleased to find that 
there was a real one, blazing away as brightly as the one she had 
left behind. `So I shall be as worm here as I was in the room,' 
thought Alice: `warmer, in fact, because there'll be no one here 
to scold me away from the fire. 
Oh, what fun it'll be, when they see me through the glass in here, 
and ca'n't get at me!'
} 
\begin{flushright}
 Lewis Carroll, {\it "Through the Looking-Glass"} 
\end{flushright}

\section{Introduction} 



Lewis Carroll's $\alice$ probably was first who seriously 
considered  that the world beyond the mirror --  
{\it "Looking-Glass House"} -- is real:   
"just same as our world, only the things go other way..." 
Observable elementary particles have left-handed $(V-A)$ 
weak interactions which violate P-parity in the strongest 
possible way.  However, there could  exist a hidden mirror 
world of particles, an exact copy of our world, 
only that mirror particles experience right-handed $(V+A)$ 
weak interactions. 
Each of ordinary particle has its mirror twin:   
"the mirror particles are something 
like our particles, only the chiralities go the wrong way..." 
Such a duplication of the worlds would restore the 
left-right symmetry of Nature, as it was suggested by  
Lee and Yang \cite{LY}. The phenomenological implications 
of such a parallel world were first addressed by  
Kobzarev, Okun and Pomeranchuk, which also introduced 
the term "Mirror World" \cite{KOP}, and several 
other papers had followed \cite{Pavsic}-\cite{FV1}.  
 
The basic concept is to have a theory given by the
product $G\times G'$ of two identical gauge groups with
the identical particle contents, which could naturally emerge
e.g. in the context of $E_8\times E'_8$ superstring. 
Once the gauge factor $G$ describes interactions of observable 
particles: quarks and leptons, Higgses, etc.,  
then its gauge counterpart $G'$ describes the world 
with analogous particle content: mirror quarks and leptons,  
mirror Higgses, etc.  
(From now on all fields and quantities of the
mirror (M) sector will be marked by $'$ 
to distinguish from the ones belonging to the
observable or ordinary (O) world.)
The M-particles are singlets of $G$ and vice versa, 
the O-particles are singlets of $G'$. 
A discrete symmetry $G\leftrightarrow G'$ interchanging
corresponding fields of $G$ and $G'$, mirror parity,
guarantees that two particle sectors have identical 
Lagrangians, with all coupling constants 
(gauge, Yukawa, Higgs) having the same pattern, and thus
their microphysics are identical.\footnote{   
Parity between two worlds can be spontaneously 
broken, e.g. the electroweak symmetry breaking 
scales in two sectors can be different, 
which would lead to somewhat different   
particle physics in mirror sector \cite{BM,BDM,BGG}. }
%
Two worlds communicate through the gravity, 
but there are also other possible ways.\footnote{  
For example, ordinary photons could have kinetic mixing with 
mirror photons \cite{Holdom,Glashow86,Glashow87},  
ordinary neutrinos could mix with mirror neutrinos \cite{FV,BM}, 
two sectors could have a common gauge symmetry of flavour 
\cite{PLB98} or common Peccei-Quinn symmetry \cite{BGG}.
} 

If the mirror sector exists, then the Universe  
along with the ordinary photons, neutrinos, baryons, etc. 
should contain their mirror partners.  
One could naively think that due to mirror parity the
O- and M- particles should have the same cosmological 
abundances and hence the two sectors should have the same
cosmological evolution. 
However, this would be in the immediate conflict 
with the Big Bang nucleosynthesis (BBN) bounds
on  the effective number of extra light neutrinos,
since the mirror photons, electrons and neutrinos
would give a contribution to the Hubble expansion rate 
equivalent to $\DN\simeq 6.14$.
Therefore, in the early Universe the M-system should have 
a lower temperature than ordinary particles. 
This situation is plausible if the following conditions  
are fulfilled:

A. After the Big Bang (post-inflationary reheating) 
the two systems get different initial temperatures, 
namely the temperature in the M-sector is lower 
than in the visible one, $T' < T$. 
This can be naturally achieved in certain models
of inflation \cite{KST,BDM,BV}.

B. The two systems interact very weakly,  
so that they do not come into thermal equilibrium
with each other during the Universe expansion.
This condition is automatically fulfilled  if the two worlds
communicate only via gravity.
If there are some other effective couplings between 
O- and M- particles, they have to be properly suppressed. 

C. Both systems expand adiabatically, without significant 
entropy production. 

If these conditions are satisfied, 
two sectors with different initial temperatures, 
evolving independently during the cosmological expansion,  
maintain the ratio of their temperatures $T'/T$ nearly 
constant at later stages.    
In this way, if $T'/T\ll 1$, 
mirror sector would not affect primordial 
nucleosynthesis in the ordinary world.  

At present, the temperature of ordinary relic photons 
is  $T\approx 2.75$ K, and the mass density of ordinary 
baryons constitutes about $5\%$ of the critical 
density. 
Mirror photons should have smaller temperature $T' < T$, 
so their number density is $n'_\ga = x^3 n_\ga$, where    
$x=T'/T$. This ratio is a key parameter in our 
further considerations as far as it remains nearly invariant 
during the expansion of the Universe. 
The BBN bound on $\DN$ implies the upper bound  
$x < 0.64\, \DN^{1/4}$. 
As for mirror baryons, {\it ad hoc} 
their number density $n'_b$  can be larger than $n_b$, 
and if the ratio $\beta = n'_b/n_b$ is about 5 or so, 
they could constitute the dark matter of the Universe.  
 
In this paper we discuss the cosmological implications 
of the mirror sector.  
We show that due to the temperature difference, in the
mirror sector all key epochs as the baryogenesis,
nucleosynthesis, etc. proceed at somewhat different
conditions than in the observable Universe.
In particular, we show that in certain baryogenesis scenarios 
the M-world gets a larger baryon asymmetry than the O-sector, 
and it is pretty plausible that $\beta > 1$ \cite{BCV}.  
This situation emerges in a particularly appealing way 
in the leptogenesis scenario due to the lepton number  
leaking from the O- to the M-sector which leads to  
$n'_b \geq n_b$, and can thus explain
the near coincidence of visible and dark components  
in a rather natural way \cite{BB-PRL}.    

Discuss the physics and the cosmology of the two worlds, 
let us also introduce two observers: 
Ordinary observer {\it Olga} and 
Mirror observer {\it Maxim}.\footnote{I do not specify here 
why I have chosen these names, but Ian Kogan would know.}  
The world of $\maxim$ is a hidden sector for $\olga$  
and vice versa, the world of $\olga$ is a hidden world 
for $\maxim$. Could they by some experimental and 
theoretical means deduce the existence of the parallel 
hidden sectors?  Also {\it Alice} can be involved 
as a super-observer which can see the whole theory, and as 
a possible messenger between two worlds.  

In the presently popular language of extra dimensions and 
brane-worlds, the concept of two paralle world can be 
visualized in a simple way. 
One could consider e.g. a five-dimensional theory 
with compactified and orbifolded fifth dimesion ($S_1/Z_2$) 
with parallel 3D-branes located in two fixed points,  
so that the ordinary matter is localized on the left-brane
$L$ and the mirror matter is localized on the right-brane $R$. 
In this view, one would simply tell that 
$\olga$ lives on the $L$-brane and  $\maxim$ on the $R$-brane, 
while $\alice$ can propagate in the bulk.

\section{Mirror world and mirror symmetry}

Let us discuss now in more details $\alice$'s theory of two 
parallel worlds. 
We consider two identical gauge factors $G\times G'$ with 
the identical representations.
Mirror parity is understood as a discrete symmetry under  
$G \to G'$, when all ordinary particles 
(fermions, Higgses and Gauge fields) exchange places with their 
mirror partners (`primed' fermions, Higgses and Gauge fields), 
so that the Lagrangian of the O-sector  
\be{Lagr-O}
{\cal L} = 
{\cal L}_{\rm Gauge} + {\cal L}_{\rm Higgs} + {\cal L}_{\rm Yuk}
\end{equation}
transforms into the Lagrangian of M-sector,  
\be{Lagr-Opr}
{\cal L}' = 
{\cal L}'_{\rm Gauge} + {\cal L}'_{\rm Higgs} + {\cal L}'_{\rm Yuk}
\end{equation}
and a whole Lagrangian ${\cal L}'+{\cal L}$ remains invariant. 

Let us consider, for simplicity, that the O-world is described 
by the Standard Model (SM) based on the gauge symmetry 
$G_{\rm SM} =SU(3)\times  SU(2)\times U(1)$, 
which has a chiral fermion content with respect to the 
electroweak gauge factor $SU(2)\times U(1)$. 
Fermions are represented as Weyl spinors, the 
left-handed (L) quarks and leptons $f_L$ transforming as doublets 
and the right-handed (R) ones $f_R$ as singlets:\footnote{
Here and in the following, we omit the family indices for 
simplicity.} 
\be{SM-L} 
f_L: ~~q_L = \dub{u_L}{d_L}, ~~ l_L = \dub{\nu_L}{e_L};  
~~~~~ 
f_R: ~~u_R, ~ d_R, ~ e_R 
\end{equation}
Then the field operators 
$\tf_R = C\ga_0 f_L^\ast$ and $\tf_L = C\ga_0 f_R^\ast$, 
$C$ being the charge conjugation matrix,  
describe antifermions which have opposite gauge charges  
as well as opposite chiralities with respect to what we call 
fermions: 
\be{SM-R} 
\tf_R: ~~ \tq_R = \dub{\tu_R}{\td_R}, ~~ 
\tl_R = \dub{\tnu_R}{\te_R}; 
~~~~~ 
\tf_L: ~~\tu_L, ~ \td_L, ~ \te_L 
\end{equation}
In addition, we prescribe a global baryon charge $B=1/3$ to 
quarks $q_L,u_R,d_R$, 
(so that baryons consisting of three quarks have $B=1$),  
and a lepton charge $L=1$ to the leptons $l_L,e_R$. 
Hence antiquarks $\tq_R,\tu_L,\td_l$ have $B=-1/3$, 
and antileptons $\tl_R,\te_L$ have $L=-1$.  
 
The Gauge and Higgs parts in the Lagrangian ${\cal L}$ are 
self-explanatory, while 
the fermion Yukawa couplings with the Higgs doublet(s) 
$\phi=\phi_{u,d}$ can be conveniently presented 
in the following form:\footnote{  
In the minimal SM, $\phi_u$ and $\phi_d$ simply are conjugated 
fields: $\phi_d \sim \phi_u^\ast$. 
However, we keep in mind that generally in the extensions of 
the SM, in particular, in the supersymmetric extension, 
$\phi_u$ and $\phi_d$ are independent ("up" and "down") Higgs 
doublets with different vacuum expectation values (VEV) 
$\langle\phi_u\rangle = v_u$ and  $\langle\phi_d\rangle = v_d$, 
and their ratio is known as $\tan\beta = v_u/v_d$.}  
\beqn{Yuk-O}
&& 
{\cal L}_{\rm Yuk} = {\cal W} + {\cal W}^\dagger \\
&&
{\cal W} = \tf_L Y f_L \phi  \equiv
\tu_L Y_u q_L \phi_u + \td_L Y_d q_L \phi_d + 
\te_L Y_e l_L \phi_d   \nonumber \\
&&
{\cal W}^\dagger = f_R Y^\ast \tf_R \tphi  \equiv
u_R Y_u^\ast \tq_R \tphi_u + d_R Y_d^\ast \tq_R \tphi_d + 
e_R Y_e^\ast \tl_R \tphi_d   
\nonumber 
\eeqn
where $Y=Y_{u,d,e}$ are the Yukawa constant matrices and 
$\tphi_{u,d} \equiv \phi_{u,d}^\ast$   
($C$-matrix and the sign of transposition are omitted for 
simplicity).  
Therefore, ${\cal W}$ is a holomorphic function of 
the L-fields and ${\cal W}^\ast$ of R-fields. 

On the other hand, the physics of M-sector,  
based on the mirror Standard Model with the gauge symmetry 
$G^\prime_{\rm SM}=SU(3)^\prime\times SU(2)^\prime\times U(1)^\prime$
has the analogous content of fermion fields:
\be{SM-Lpr} 
f^\prime_L: ~~q^\prime_L = \dub{u'_L}{d'_L}, ~~ 
l^\prime_L = \dub{\nu'_L}{e'_L};  
~~~~~ 
f^\prime_R: ~~u'_R, ~ d'_R, ~ e'_R, 
\end{equation}
and that of anti-fermions 
\be{SM-Rpr} 
\tf^\prime_R: ~~ \tq'_R = \dub{\tu'_R}{\td'_R}, ~~ 
\tl^\prime_R = \dub{\tnu'_R}{\te'_R}; 
~~~~~ 
\tf_L: ~~\tu'_L, ~ \td'_L, ~ \te'_L, 
\end{equation}
For definiteness, let us precribe mirror fermion numbers: 
$B'=1/3$ to quarks $q'_L,u'_R,d'_R$ and $L'=1$ 
and leptons $l'_L,e'_R$, 
and so antiquarks $\tq'_R,\tu'_L,\td'_l$ have $B'=-1/3$, 
and antileptons $\tl'_R,\te_L$ have $L'=-1$.  

Yukawa coupligs have the form analogous to (\ref{Yuk-O}): 
\beqn{Yuk-M}
&& 
{\cal L}^\prime_{\rm Yuk} = 
{\cal W}^\prime + {\cal W}^{\prime\dagger} \\
&&
{\cal W}^\prime = \tf^{\prime}_L Y' f'_L \phi' \equiv
\tu^{\prime}_L Y'_u q'_L \phi'_u + 
\td^{\prime}_L Y'_d q'_L \phi'_d + 
\te^{\prime}_L Y'_e l'_L \phi'_d    \nonumber \\
&&
{\cal W}^{\prime\dagger} = 
f^{\prime}_R Y^{\prime\ast} \tf'_R\tphi' \equiv 
u^{\prime}_R Y_u^{\prime\ast} \tq'_R \tphi'_u + 
d^{\prime}_R Y_d^{\prime\ast} \tq'_R \tphi'_d + 
e^{\prime}_R Y_e^{\prime\ast} \tl'_R \tphi'_d   
\nonumber 
\eeqn
where $\phi'=\phi'_{u,d}$ are the mirror Higgs doublets. 

What kind of discrete symmetries can have such a theory? 

Let us consider first O-sector separately. 
The weak interactions of ordinary particles 
break one of the possible fundamental symmetries 
of the Nature, parity, in a strongest possible way.  
The physics is not invariant under the coordinate transformation 
$x \to -x$, and hence the left- and right-handed systems 
of the coordinates are not equivalent.  

Namely, for what she calls particles: baryons and leptons,
(probably because she herself is made up of them),    
the weak interactions have the left-handed ($V-A$) form.

The particle content of the Standard Model and hence its 
Lagrangian is not symmetric under the exchange of the 
L and  R particles: $f_L \tolr f_R$.  
In particular, the gauge bosons of $SU(2)$ couple to the 
$f_L$ fields but do not couple to $f_R$ ones.   
In fact, in the limit of unbroken $SU(2)\times U(1)$ symmetry, 
$f_L$ and $f_R$ are essentially independent species 
with different quantum charges. 
The only reason why we call e.g. two Weyl fermions 
$e_L\subset l_L$ and $e_R$ respectively 
as the left- and right-handed electrons is that 
after the electroweak breaking down to $U(1)_{\rm em}$ 
these two have the same electric charges and form a massive 
Dirac fermion $\psi_e = e_L+e_R$. 

There exist left-right extensions of the Standard Model, 
with the electroweak gauge symmetry extended to
$SU(2)_L\times SU(2)_R\times U(1)_{B-L}$. 
Now $f_L$ are doublets of $SU(2)_L$ and $f_R$ (including 
right-handed neutrinos) are doublets of $SU(2)_R$. 
Lagrangian of such a theory can be invariant under the 
exchange $f_L\to f_R$ if at the same time two gauge 
sectors interchange the places: $SU(2)_L \to SU(2)_R$. 
However, experiment tells that if such a parity exists, 
than it should be spontaneously broken:  
mass of the $W_R$ gauge bosons should be much larger than 
the mass of ordinary $W_L$ bosons.  
 
Once again, for what we call particles $f$ (\ref{SM-L}), 
their weak interactions are left-handed ($V-A$),  
while in terms of antiparticles $\tf$ (\ref{SM-R}), 
the weak interactions would be seen as right-handed ($V+A$), 
since now only R states couple to the $SU(2)$ bosons. 
In the context of the SM or its extensions, 
the symmetry between particles $f$ and antiparticles $\tf$, CP-parity,  
could be the the only exact fundamental symmetry between the 
left and right: $f_L\to \tf_R$, ($f_R \to \tf_L$), 
as far as it is respected by the gauge interactions. 
But not by the Yukawa terms!
Although the terms (\ref{Yuk-O}) are presented 
in an rather symmetric manner between $L$ and 
$\tR$ fields, ${\cal W}$ being a holomorphic function of only 
$L$ fields and ${\cal W}^\ast$ of $\tilde R$ fields,   
they are not invariant under the interchange $L\to \tR$ due 
to irremovable complex phases in the Yukawa coupling matrices 
$Y_{u,d,e}$. As we know, after the electroweak symmetry breaking, 
the complexity propagates to the CKM matrix of the quark mixing and 
leads to observable CP-violation effects. 
Hence, neither CP-parity is respected. 
Nature once again demonstrates that 
"the only good parity is a broken parity".  

Clearly, one could always redefine the notion of particles 
and antiparticles, to rename particles as antiparticles and 
vice versa. 
The natural choice for what to call particles 
is given by the content of matter in our Universe.  
Matter, at least in our galaxy and its wide neighbourhoods, 
consists of baryons $q$ while 
antibaryons $\tq$ can be met only in accelerators or perhaps 
in cosmic rays. However, if by chance we would live in the 
antibaryonic island of the Universe, we would claim that 
our weak interactions are right-handed.  
Therefore, not only our microphysics 
but also our Universe does not 
respect the exchange between matter and antimatter. 

The CP-violating effects in particle physics are tiny. 
In fact, most of the elementary processes are exactly 
the same between the particles and antiparticles, 
and CP-violation is observed only in some rare processes, 
as e.g. leptonic asymmetry in $K_L$ decay. 

On the other had, our Universe, presently being completely 
dominated by matter over antimatter, 
$n_b/n_{\bar b}\gg 10^{10}$, 
at the early stages was almost symmetric, with a tiny excess of 
baryons over antibaryons 
$n_b/n_{\bar b} = 1+ {\cal O}(10^{-10})$. 
It is a very profound question, who and 
how has prepared our Universe at the initial state  
to provide a tiny excess of baryons over antibaryons, 
and therefore fixed a priority of the $V-A$ form 
of the weak interactions over the $V+A$ one. 
It is appealing to think that the 
baryon asymmetry itself emerges due to the tiny  
CP-violating features in the particle interactions, 
and it is related to some fundamental physics beyond the 
Standard Model which is responsible for the primordial 
baryogenesis.  
Without the CP-violating effects, probably, 
no Baryon Asymmetry would be possible at all, 
and the Universe would consist only of light. 

Concluding, the particle physics of ordinary world is not 
symmetric between the left and right -- 
neither $P:$ $f_L \to f_R$  nor 
$CP:$ $f_L \to \tf_R$ are exact symmetries --  
they are explicitly (or perhaps spontaneously) broken. 

However, the whole theory describing both worlds, 
can be symmetric between left and right. 
Indeed, consider that transformation of coordinates 
P: $x\to -x$ is accompanied by the transformation which 
interchanges O-fields $f,\phi$ with the corresponding M-fields $\tf,\tphi$ 
in the following manner \cite{FV1}:
\be{MLR}
f_L,\tf_L \to \gamma_0 \tf'_R,\gamma_0f'_R,  ~~~~ 
\tf_R, f_R \to \gamma_0 f'_L,\gamma_0\tf'_L, 
~~~~ \phi \to \tphi' 
\end{equation}
while also the gauge fields of $G$ properly transform 
in their partners in $G'$. Such a symmetry can be called 
matter parity $MP$.
Obvioulsy, it implies that the gauge couplings are exactly 
the same two sectors, the Higgs potentials are identical, 
while for the Yukawa coupling constants we have 
\be{MLR-Yuk} 
Y'_{u,d,e} = Y^\ast_{u,d,e}
\end{equation}
In this way, the introducing of M-world does not introduce  
any new parameter 
So, in this case the particle physics of the M-world 
will be exactly the same in terms of the R-fields $\tf'_R,f'_R$ 
as that of the O-world in terms of L-fields $f_L,\tf_L$.
Hence, $MP$ restores the left-right symmetry as a symmetry 
between two sectors. 

The generalization for the extensions of the Standard Model 
related to supersymmetry and grand unification is strightforward.
Consider a generic supersymetric gauge theory with $G\times G'$
symmetry where the O-sector is presented by a set 
left-chiral superfields $L$ in certain representations of $G$ 
and M-sector by set of left-chiral superfields $\tilde L$ 
in analogous (anti)representations of $G'$, so that $L$ are 
singlets of $G'$ and vice versa, $\tilde L$ are singlets of $G$. 
One can also explicitly write thei conjugated 
right-chiral superfields 
$\tR = L^\ast$, $\tR'=L^{\prime\ast}$ 

For example, in the context of the MSSM, both 
O-matter and M-matter are presented in terms 
of left-chiral superfields $L$ and their conjugated 
right-chiral superfields $\tR$:\footnote{
In the context of $N=1$ supersymmetry, fermions and  
Higgses become chiral superfields, and formally 
they can be distinguished only by matter parity. 
} 
\beqn{LL-SM}
&&
L: ~~ q,l,\tu,\td,\te, ~\phi_{u,d} \, ; ~~~~~~~~ 
\tR: ~~ \tq,\tl,u,d,e, ~ \tphi_{u,d}  \nonumber \\
&&
L': ~~ q',l',\tu',\td',\te', ~\phi'_{u,d} \, ; 
~~~ 
\tR': ~ \tq',\tl',u',d',e', ~ \tphi'_{u,d}  
\eeqn
with the superpotential terms  
\beqn{W}
&&
W = \tu Y_u q \phi_u + \td Y_d q \phi_d + \te Y_e l \phi_d ~+~ 
\mu \phi_u\phi_d  \nonumber \\ 
&&
W^\prime = \tu^{\prime} Y'_u q' \phi'_u + 
\td^{\prime } Y'_d q' \phi'_d + 
\te^{\prime } Y'_e l' \phi'_d +    
\mu' \phi'_u\phi'_d  
\eeqn  
In $SU(5)\times SU(5)'$ model  
$L$ should include the fermion superfields $\bar5+10$ and 
the Higgs superfields $5,\bar5$ and 24 of $SU(5)$, and 
$\tilde L$ the same superfields of $SU(5)'$. 

Then the Lagrangian reads: 
\beqn{susy-W}
&& 
\cL_{\rm mat} =  \int d^2\theta W(L) ~ + ~ 
\int d^2\ov{\theta} W^\ast(\tR) ~ + ~ 
\cL_{\rm Gauge}  \nonumber \\ 
&&
\cL'_{\rm mat} =  \int d^2\theta W'(L') ~ + ~ 
\int d^2\ov{\theta} W^{\prime\ast}(\tR') ~ + ~ 
\cL'_{\rm Gauge}   
\eeqn 
with superpotentials $W(L)$ and $W'(L')$ 
being holomorphic functions respectively of the superfields 
$L$ and $\tL$.   

Then, M-parity can be understood as transformation 
of all left superfields $L$ of $G$ into corresponding 
right superfields $\tR'$ of $G'$, 
$L\to \tR'$ and $L' \to \tR$, accompanied by appropriate 
exchange between the vector superfields of $G$ and $G'$. 
and hence it implies that  
$W$ and $W{\prime\ast}$ have the same functional shape, 
or $W$ and $W'$ are complex conjugated. Therefore, all complex 
coupling constants in $W'$ should have the opposite phase 
with respect the corresponding ones in $W$.    
In particular, for the superpotential (\ref{susy-W}) 
MP implies that 
$Y'_{u,d,e} = Y^\ast_{u,d,e}$ and $\mu'=\mu^\ast$.

Concluding, the $MP$ implies that 
M-world has the same physics in terms of the R-states 
as the ordinary one in terms of the L-states. 
However, this does not mean that macroscopic realisations 
of M-worlds would necessarily be the mirror reflection 
of O-world. 
The sign of baryon asymmetry is crucial for 
determination of the nature of M-world.  
 
Let us discuss now how the particle physics of two 
worlds can be seen by ordinary and mirror observers 
in different situations. 

In ordinary world baryons dominate over antibaryons, 
i.e. $B>0$.    
In this view, the ordinary observer $\olga$ 
identifies $f$ species as particles (matter) and 
$\tf$-species as antiparticles (antimatter).  
His experimental results can be then formulated as: 

$\bullet$ 
$P$ is broken. 
Matter of O-world has left-handed nature: 
weak interactions of O-particles have $V-A$ form, 
neutrinos are L-handed.  

$\bullet$ $CP$ is broken.  
Decays of $K_L$ meson demonstrate a tiny excess 
of the positrons $e^+$ over the electrons $e^-$.  

As for the M-sector, two different situations are possible. 
The experimental results of the mirror observer $\maxim$ 
would depend on the sign of baryon asymmetry in M-world. 
Apriori it can be either positive or negative. 

Namely, 
if in M-world $\tf'$ species dominate over $f'$, i.e. $B' < 0$,    
then $\maxim$ would identify the former as particles and 
the latter as antiparticles. In this case his conclusions 
would be: 

$\bullet$ 
$P$ is broken. 
Matter of M-world has right-handed nature: 
weak interactions of O-particles have $V+A$ form, 
neutrinos are R-handed. 

$\bullet$ $CP$ is broken.  
Decays of $K_L$ meson demonstrate a tiny excess 
of the positrons $e^+$ over the electrons $e^-$.  

Obviously, this situation will be necessarily realised 
if baryon asymmetries in two sectors are induced 
separately, by the same particle physics mechanism 
related to $CP$-violating phases which are opposite 
for the Yukawa couplings of $L$ and $L'$.   

However, it is also possible that 
M-world is dominated by $f'$ species over $\tf'$, 
i.e. $B' > 0$.     
In this case naturally 
$\maxim$ would identify particles as $f'$ and 
antiparticles as $\tf'$, and thus he would conclude that:  

$\bullet$ 
$P$ is broken. 
Matter of O-world has left-handed nature: 
weak interactions of O-particles have $V-A$ form, 
neutrinos are L-handed.  

$\bullet$ $CP$ is broken.  
Decays of $K_L$ meson demonstrate a tiny excess 
of the electrons $e^-$ over the positrons $e^+$.  

As we shall see later, such a situation can be realized if 
baryon asymmetries in both worlds arise by unique 
mechanism related to the particle interactions between two 
worlds via some messengers as gauge singlet right-handed 
neutrinos.  

Alternatively, one could impose between two sectors another 
type of parity, $D$-symmetry under transformation 
$f_{L,R} \tolr f'_{L,R}$, 
$\phi \tolr \phi'$, which instead of (\ref{MLR-Yuk}) 
would imply $Y'_{u,d,e} = Y_{u,d,e}$. 
This case is nothing but direct duplication. 
Obviously, if both $MP$ and $D$  are 
imposed, the whole theory would be CP-invariant. 
  
Either type of parity implies that the two sectors 
have the same particle physics.
If the two sectors are separate and do not interact by forces 
other than gravity, the difference between $D$ and $MP$  
is rather symbolic and does not have any profound implications. 
However, in scenarios with some particle messengers 
between the two sectors this difference can be important 
and can have dynamical consequences.  

It is not clear how two observers could communicate 
to each other the information about microscopic nature of 
their worlds -- can they encode it e.g. into polarized 
gravitational waves or even photons? 
(if there is a kinetic mixing between two photons).  
Perhaps $\alice$, once again, could play the role of the 
messenger: in fact, 
 Ian Kogan with collaborators have considered  
Alice string in extra dimension as a possible passage 
from one parallel sector to another \cite{DKS}. 

\section{Spontaneously broken Mirror Parity} 

There is no reason to expect that Nature does not apply 
the old principle "The only good parity ... is a broken parity". 
It is interesting to discuss under which circumstances 
$MP$ could be broken. 

If O- and M-sectors are described by the minimal Standard Model, 
with the Higgs doublets $\phi$ and $\phi'$ having identical 
Higgs potentials: 
\beqn{V-Vpr}
&& 
{\cal V}_{\rm Higgs} = 
-\mu^2 \phi^\dagger\phi + \lambda^2(\phi^\dagger\phi)^2    
\nonumber \\
&&
{\cal V}'_{\rm Higgs} = 
-\mu^2 \phi^{\prime\dagger}\phi' + 
\lambda^2(\phi^{\prime\dagger}\phi')^2    
\eeqn 
then the VEVs of $\phi$ and $\phi'$ are unavoidably 
identical: $v=v'=\mu/\lambda$. 
In addition, from 
the experimental limits on the Higgs mass one can 
conclude that $\lambda \sim 1$. 

The gauge symmetry of the theory allows also 
a quartic interaction term between O-and M-Higgs doublets:  
\be{HHpr} 
{\cal V}_{\rm mix} = 
\kappa (\phi^\dagger \phi)(\phi^{\prime\dagger}\phi') 
\end{equation}
This term is cosmologically dangerous, since it would bring 
the two sectors into equilibrium in the early Universe via 
interactions $\bar{\phi}\phi \to \bar{\phi}'\phi'$ 
unless $\kappa$ is very small, $\kappa < 10^{-8}$ \cite{BDM}. 
As far as $\kappa \ll \lambda$, this mixed term cannot cause 
the asymmetry between $v$ and $v'$. 

For achieving the breaking of mirror parity, one has to extend 
the Higgs sector. The simplest way is to introduce a real singlet 
scalar $\eta$ which is odd under under mirror parity: 
it changes the sign when two gauge sector exchange the places: 
$\eta \to - \eta$. Therefore, its interaction Lagrangian 
with the O- and M-Higgs doublets should include terms 
$\propto \eta(\phi^\dagger \phi - \phi^{\prime\dagger}\phi')$. 
Therefore, if $\eta$ has a non-zero VEV, it would induce 
difference in mass-squared terms of $\phi$ and $\phi'$ 
and hence mirror weak scale can be different from the 
ordinary one \cite{BDM}. But such an extension of the Higgs sector 
is not the most beautiful thing that one can imagine.   

The situation remains the same if both O- and M-sectors 
are described by MSSM. 
In this case ordinary Higgses $\phi_{u,d}$  
as well as their mirror partners $\phi'_{u,d}$ 
become chiral superfields and thus their renormalizable 
Lagrangian can include no mixed terms. Namely, the F-terms 
read as   
\be{susyHH} 
{\cal L}_{\rm Higgs} =
\int d^2\theta (\mu\phi_u\phi_d + \mu\phi'_u\phi'_d) 
~ + ~ {\rm h.c.}
\end{equation}
and clearly also gauge D-terms of O- and M-sectors 
are unmixed. In addition, experimental limits 
imply that $\mu$-terms are of the order 
of 100 GeV.  

The minimal gauge invariant term between the 
O- and M-Higgses in the superpotential has dimension 5: 
$(1/M)(\phi_u\phi_d) (\phi'_u\phi'_d)$, 
where $M$ is some large cutoff mass, e.g. of the order 
of the GUT or Planck scale.     
If this term is inculded together with (\ref{susyHH}), 
a mixed quartic terms similar to (\ref{HHpr}) emerge in the 
Lagrangian: 
\be{HH}
\la (\phi_u^\dagger\phi_u)(\phi'_u\phi'_d) + 
\la (\phi_d^\dagger\phi_d)(\phi'_u\phi'_d) + 
(\phi_{u,d}\to \phi'_{u,d}) + {\rm h.c.} 
\end{equation}
with the coupling constant $\la = \mu/M \ll 1$.

Neither the soft supersymmetry breaking $F$-term and $D$-terms 
are dangerous. For example, the F-term 
$\frac{1}{M}\int d^2\theta z (\phi_u \phi_d)(\phi'_u \phi'_d) + $ h.c.,  
where $z=m_S\theta^2$ being the supersymmetry breaking spurion,
gives rise to a quartic scalar term 
\be{HH1}
\la (\phi_u\phi_d)(\phi'_u\phi'_d) + {\rm h.c.} 
\end{equation}
with $\la \sim m_S/M \ll 1$. Thus for $\mu,m_S\sim 100$ GeV, 
all these quartic constants are strongly suppressed, and 
hence are safe. 

It is easy, however, to construct the simple supersymetric 
model where mirror parity is spontaneously broken. 
Let us introduce an additional singlet superfield
$S$ and consider the Higgs superpotential having the form: 
\be{W-Higgs}
W = \lambda S (\phi_u\phi_d + \phi'_u\phi'_d - {\cal M}^2) + 
MS^2 + ... 
\end{equation}
where $M$ and ${\cal M}$ are 
some large mass scales, much larger than $M_W$.  
It is simple to see that such a theory can bring to the 
spontaneous breaking of M-parity due to different VEVs 
of the O- and M- Higgs doublets. Namely, mirror Higgses 
can get VEVs $v'_u=v'_d={\cal M}$ while the ordinary ones  
get VEVs of the order of the supersymmetry breaking scale. 
The hierarchy problem between two VEVs is solved via the 
so called GIFT (Goldstones Instead of Fine Tuning) 
mechansim, which functions  
when the superpotential has an accidental global symmetry 
bigger than the local symmetry of the theory \cite{GIFT}.  
Nemaly, the Higgs doublets emerge as pseudo-Goldstone 
modes of the accidental global $SU(4)$ symmetry possesed 
by superpotential terms (\ref{W-Higgs}). In this way, 
the Mirror world can appear useful for solving the ordinary 
theoretical problems in ordinary world, as is the problem 
of the Higgs mass/VEV stability without fine tuning of
parameters.\footnote{Let me mention also another interesting 
example, when the presence of mirror sector can be used 
for the naturallness with respect of the flavour-changing 
problem \cite{PLB98}.    
}

Let us first consider the minimum of such a theory in the 
supersymmetric limit, without taking into account the 
soft supersymmetry breaking terms (soft masses and trilinear 
$A$-terms).  
Then the F-term condition $F_S=0$ tells that 
$v_uv_d + v'_uv'_d = {\cal M}^2$, but does not fixes to how 
the VEVs are distributed between $\phi_{u,d}$ and $\phi'_{u,d}$
Higgses. In fact, this superpotential has an accidental global 
symmetry $U(4)$, larger than the local symmetry $U(2)\times U(2)'$. 
Two "upper" doublets ($\phi_u$,$\phi'_u$) "down" doublets 
as $(\phi_d,\phi'_d)$, form 4-plets of $SU(4)$.  
So if one choses that $v_{u,d}=0$ in the supersymmetric limit, 
than we have $v'_uv'_d = {\cal M}^2$, and in addition from 
the  D-term condition of $SU(2)'$ it follows that 
$v'_u=v'_d=v'/\sqrt2$, i.e. $\tan\beta'=1$. So the mirror 
standard model is broken at the scale $\sim {\cal M}$ 
but the ordinary one remains unbroken. 
As for the doublets $\phi_{u,d}$, yet in the limit of unbroken 
supersymmetry they remain as massless Goldstone modes of 
the spontaneously broken accidental symmetry $SU(4)$, and their 
mass terms/VEVs can emerge only after the supersymmetry breaking. 
At the first approximation, also sift terms have an accidental 
global symnmetry $SU(4)$ (in particular, $A$ terms repeat 
the structure of the superpotential (\ref{W-Higgs}), 
therefore one combination of two Higgses $\phi_u$ and $\phi_d$ 
gets a mass $\sim m_S$ but another remains as a Goldstone mode. 
(In adition, the non-zero $\mu$-term is generated for 
since $S$ will get a VEV of the order of $m_S$.) The mass 
and VEV of the latter will emerge only after accounting for 
the terms which explicitly break the accidental $SU(4)$, 
which are the MSSM $D$-terms and the Yukawa couplings 
among which the relevant one is $\lambda_t$.  

Let us discuss briefly how would look the particle physics 
of the mirror sector if $MP$ is spontaneously broken 
(for more details, see \cite{BDM}). 

In the Standard Model, with one Higgs $\phi$, 
the electroweak breaking scale identified with the Higgs VEV, 
is unambiguously fixed by the Fermi constant:  
$v=174$ GeV, and we have to seriously take this into account. 
In the models with two Higgs doublets, $\phi_u$ and $\phi_d$, 
the VEVs $v_u$ and $v_d$ cannot be known separately, 
the parameter $\tan\beta =v_u/v_d$ can be arbitrary 
(well, with all probability larger than 1 but smaller than 100), 
however the total VEV $v_u^2 + v_d^2 = v^2$.   

Let us take now the mirror electroweak scale
$\langle \phi' \rangle =v'$ different from $\langle \phi \rangle =v$.
Namely, let us assume that $v'\gg v$ 
(for the moment, we assume for a simplicity 
that $\tan\beta'=\tan\beta$ 
once the supersymmetric model is concerned).  
As far as the Yukawa couplings have the same values in both systems,
the mass and mixing pattern of the charged fermions in the mirror world
is completely analogous to that of the visible one, but with all
fermion masses scaled up by the factor $\zeta=v'/v$ 
The masses of gauge bosons and higgses are also scaled as
$M_{W',Z',\phi'}=\zeta M_{W,Z,\phi}$
while photons and gluons remain massless in both sectors.

With regard to the two chromodynamics,
a big difference between the electroweak scales $v'$ and $v$   
will not cause the similar big
difference between the confinement scales
in two worlds. Indeed, if $P$ parity is valid at higher (GUT) scales,
the strong coupling constants in both sectors would evolve down in energy
with same values until the energy
reaches the value of the mirror-top ($t'$) mass.
Below it $\alpha'_{s}$ will have a different slope than $\alpha_{s}$.  
It is then very easy to calculate the value of the scale $\Lambda'$
at which $\alpha'_{s}$ becomes large.
This value of course depends on the ratio $\zeta= v'/v$. 

Taking $\Lambda=200$ MeV for the ordinary QCD, then for 
e.g. $\zeta\sim 30$ we find $\Lambda'\simeq 300$ MeV or so.
On the other hand, we have $m'_{u,d}=\zeta m_{u,d}\sim m_s$ so that
masses of the mirror light quarks $u'$ and $d'$
do not exceed $\Lambda'$. So the condensates
$\langle \bar{q}' q' \rangle$ should be formed with
approximately the same magnitudes as
the usual quark condensates $\langle \bar{q}q \rangle$.
As a result, mirror pions should have mass
$m'_{\pi}\simeq \sqrt{(m_{u'}+m_{d'})\langle \bar q' q'\rangle }$
comparable to the mass of normal Kaons
$m_{K}\simeq \sqrt{m_{s}\langle \bar q q \rangle }$.    
   
As for the mirror nucleons,
their masses are approximately 1.5 times larger
than that of the usual nucleons.
Since $(m'_d-m'_u)\approx 30 (m_d-m_u)$
we expect the mirror neutron $n'$ to be heavier than the mirror
proton $p'$ by about 150 MeV or so,
while the mirror electron mass is $m'_e=\zeta m_e\sim 15$ MeV.
Clearly, such a large mass difference cannot be compensated by
the nuclear binding energy and hence
even bound neutrons will be unstable
against $\beta$ decay $n'\to p' e' \bar{\nu}'_e$.
Thus in the mirror world
hydrogen will be the only stable nucleus.

\begin{figure}[ht]
\centerline{\epsfxsize=3.9in\epsfbox{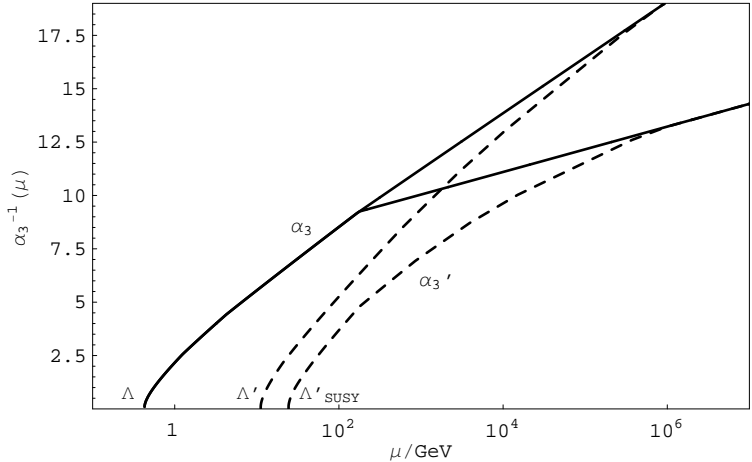}}
\caption{
Evolution of strong coupling constant in two sectors from 
high energies to lower energies in the case of $v'=10^6$ GeV 
for the Standard Model case and the case of MSSM 
with Supersymmetry broken at 200 GEV (upper an lower respectively).   
Solid curve stands for O-sector and dashed curve for M-sector. 
\label{ian-lamb}}
\end{figure}

Certainly, for bigger $\zeta$, $\Lambda'/\Lambda$ increases further. 
Also, the increasing of $\Lambda'$ with respect to $\Lambda$ 
is stronger in supersymmetric model,  and it 
can easily reach few GeV for $v' \sim 10^5-10^6$ GeV 
(see Fig. 1). This can have important impact on the dark 
matter properties if the latter is constituted by mirror 
baryons: first, that mirror baryons now can have an order 
of magnitude bigger mass then the ordinary ones, and thus 
explain to why $\Omega'_B \sim 5 \Omega_B$ under the 
situation $n'_B=n_B$ which, it turn, can naturally occur 
in the lepto-baryogenesis model \cite{BB-PRL}. Second, 
the mirror electron mass increases by 3-4 orders of magnitude 
and correspondingly decreases the radius of the mirror hydrogen 
atom with respect to that of the ordinary one. Therefore, 
in this case the mirror matter becomes weakly collisional 
and non-disipative.  

In the context of the supersymmetric theory (or more 
generically of the two Higgs-doublet models), also 
$\tan\beta'$ can be different from $\tan\beta$. 
In this case, even for $v' \sim v$, one could have an interesting 
situation when the mirror neutron becomes lighter than the 
mirror proton, and so M-proton becomes instable with respect 
to $\beta$-decay into M-neutron. 
Then that the mirror baryon (dark) matter 
is essentially constituted by M-neutrons, which is another 
interesting weakly collisional and non-dissipative sort 
of dark matter.   

Concluding, if the mirror parity is broken, then the 
microphysics of mirror world would not be the same 
as that of the ordinary world, and M-sector essentially becomes 
a sort of hidden sector with particle properties which only 
can be guessed (deduced) to some extend from the coupling constant 
structures of ordinary world.  
It is also a good exercise for thinking of anthropic/environmental 
principles, for  understanding what could happen 
to our world should the electroweak scale be different. 

Nevertheless, in further considerations we mostly concentrate 
on the case of the exact mirror parity which implies the 
same microphysics for both ordinary and mirror worlds.

\section{ Neutrinos as messengers between O- and M-worlds } 


In the context of the minimal Standard Model neutrinos can get 
the Majorana masses from the dimension 5 operator  
cutoff by large mass scale $M$: 
\be{op-nu}
\frac{1}{2 M} (\phi l)A(l\phi) +  {\rm h.c.} 
\end{equation}
where $A$ is a symmetric $3\times 3$ matrix 
of the coupling constants and $\phi\equiv \phi_u$.
After substituting the Higgs VEV $\langle\phi \rangle = v$,
one obtains the neutrino mass matrix   
$\hat{m}_\nu = A \cdot v^2/M$. 
This naturally explains why neutrinos are much lighter 
than the charged leptons and quarks. 
The latter are Dirac fermions with masses proportional to the 
weak scale $v$, whereas the neutrinos are Majorana fermions with 
masses $\sim v^2/M$.


Similar operator of dimension 5 
\be{op-nupr}
\frac{1}{2 M} (\phi'l')A'(l'\phi') + {\rm h.c.} 
\end{equation}
induces the Majorana mass matrix of M-neutrinos,  
$\hat{m}'_\nu = A' \cdot v^{\prime 2}/M$, via the M-Higgs VEV 
$\langle\phi^\prime \rangle = v^\prime$.\footnote{
Although in this paper we mostly concentrate 
on the case with exact mirror parity, one should admite a 
possibility that MP  could be spontaneously broken e.g. 
so that the weak interaction scales
$\langle \phi \rangle =v$ and $\langle \phi' \rangle =v'$
are different (see next section). 
This leads to somewhat different particle
physics in the mirror sector. 
The models with spontaneoulsy  
broken parity and their implications were considered in 
refs. \cite{BM,BDM,BV}. 
}

However, the mixed operator of dimension 5 is also allowed:  
\be{op-numix} 
\frac{1}{M} \phi l D l'\phi' + {\rm h.c.} 
\end{equation}
After substituting the Higgs VEVs it leads to mixed
mass matrix $\hat{m}_{\nu\nu^\prime} = D \cdot v v^\prime/M$. 
with $D$ being $3\times 3$ matrix of the coupling constants. 
Thus, the total $6\times 6$ mass matrix of ordinary neutrinos 
$\nu\subset l$ and their mirror partners 
$\nu^\prime\subset l^\prime$ reads as 
\cite{BM}:
\be{numass}
\mat{\hat{m}_\nu}{\hat{m}_{\nu\nu^\prime}}
{\hat{m}_{\nu\nu^\prime}^T}{\hat{m}_{\nu^\prime}} =
\frac{v^2}{M} \mat{A}{\zeta D}{\zeta D^T}
{\zeta^2 A^\prime } \, .
\end{equation}
where $\zeta=v'/v$. 
Mirror Parity (\ref{MLR}) imposes the following constraints 
on the coupling constant matrices 
\be{AApr}
A' = A^\ast,  ~~~~ D=D^\dagger 
\end{equation}

In general, this matrix describes six mass eigenstates 
of Majorana neutrinos, which are superpositions of 
three ordinary neutrinos  and three mirror neutrinos. 
In the language of neutrino physicists, the O-neutrinos 
$\nu_e,\nu_\mu,\nu_\tau$ are {\it active} neutrinos while 
the M-neutrinos $\nu'_e,\nu'_\mu,\nu_\tau$ are 
{\it sterile} neutrinos.  
Thus, this model provides a simple explanation of
why sterile neutrinos could be light
(on the same grounds as the active neutrinos)
and could have significant mixing with the active neutrinos.

If parity between two worlds is exact, $\zeta=1$, then neutrinos 
of two sectors will be strongly mixed. It seems difficult 
to reconcile this situation with the present experimental 
and cosmological limits on the active-sterile neutrino mixing,  
however it is still premature to conclude that it is ruled out. 
If instead mirror parity is spontaneously broken, so that 
e.g. $\zeta=v'/v\gg 1$, then the active-sterile mixing angles should be 
small: $\theta_{\nu\nu^\prime}\sim 1/\zeta$, while their mass ratios 
(per each flavour) scale as $m_\nu/m_{\nu^\prime} \sim 1/\zeta^2 \gg 1$. 

The situation when the O- and M-neutrinos have separate 
mass terms and 
thus there are no active-sterile neutrino oscillations 
corresponds to a particular case $D=0$, and in general 
it should be motivated by some additional symmetry reasons. 

Another interesting case corresponds to $A,A'=0$ but $D\neq 0$. 
It can emerge if two sectors have a common lepton number 
(or $B-L$) symmetry: ordinary leptons $l$ have lepton charges 
$L=1$ and mirror ones $l'$ have $L=-1$. 
Obviously, this symmetry would forbid the terms  
(\ref{op-nu}) and (\ref{op-nupr}),        
while the operator (\ref{op-numix}) is allowed.  
Then, `Majorana' mass terms are absent both for 
O- and M-neutrinos in (\ref{numass}) and neutrinos are the 
Dirac fermions with {\it naturally small} masses $\sim vv'/M$, 
having left components  
$\nu_L\subset l$ in O-sector and right components  
$\tnu'_R\subset \tl$ in M-sector.    

Let us consider the situation in the context of the 
seesaw mechanism is nothing but a natural way of generating
the effective operators (\ref{op-nu}) etc. 
from the renormalizable couplings.  
In the context of the Standard Model, in addition to the  
fermions (\ref{SM-L}) with non-zero gauge charges, 
one introduces also the gauge singlets,  
so called right-handed neutrinos $\tN_a=\tN_{kR}$  
(or their conjugated left-handed states 
$N_a = C\gamma_0 \tN_a^\ast$), 
with the large Majorana mass terms 
$\frac12 (M_{ab} N_a N_b + M_{ab}^\ast \tN_a \tN_b)$. 
The mass matrix $M$ is symmetric with respect to indices 
$a,b=1,2,...,n$, and it is convenient   
to parametrize it as $M_{ab} = g_{ab}M$, 
where $M$ is a typical mass scale and $g$ is a matrix of 
dimensionless Yukawa-like constants.\footnote{
Notice, that the number of heavy singlet neutrinos $n$ does not 
have to coincide with the number of standard families $n_g=3$.
From phenomenological constraints, $n=2$ or $n>3$ are also 
possible. In this view, generically $g$ is a symmetric  
$n\times n$ matrix while $y$ and $y'$ are 
$3\times n$ matrices. } 
The ordinary leptons $l$ can couple to $N$   
via Yukawa terms analogous to (\ref{Yuk-O}): 
$y_{ia}l_i N_a \phi + {\rm h.c.}$ 
However, for $N$ being the gauge singlets, 
the mirror leptons $l'$ can couple with them with the same 
rights:  $y'_{ia}l'_i N_a \phi' + {\rm h.c.}$.  
In this way, $N$ play the role of messengers between ordinary
and mirror particles.  
The whole set of relevant terms in the Lagrangian has the form:
\be{Yuk-nu}
y_{ia}l_i N_a \phi + y^\prime_{ia}\lpr_i N_a \phpr +
\frac{M}{2} g_{ab} N_a N_b
+  {\rm h.c.}
\end{equation}
After integrating out the heavy states $N$, all 
operators (\ref{op-nu}), (\ref{op-nupr}) and (\ref{op-numix}) 
are induced with the coupling constants 
\be{AAD} 
A=yg^{-1}y^T, ~~~~ A'=y'g^{-1}y^{\prime T}, ~~~~ 
D=yg^{-1}y^{\prime T}
\end{equation} 
Without loss of generality, the matrix $G$ can be taken 
real and diagonal. Then, assuming for simplicity that 
all states $N_a$ have positive mirror parity, 
$MP:$ $N_a \to N_a$, when $l_i \to \tl_i$, 
we see that $MP$ requires that $y'_{ia}=y_{ia}^\ast$, 
and so for the coupling constants of the effective operators 
we obtain constraints (\ref{AApr}).\footnote{In general, 
some of $N_a$ can have positive and others negative parity, 
i.e. $N_a \to p_a N_a$ where $p_a = \pm 1$. This would lead 
to $y'_{ia} = p_a y_{ia}^\ast$. }  

In the next sections we show that the $N$ 
states can mediate L and CP violating scattering processes 
between the O- and M-sectors which could provide a new  
mechanism for primordial leptogenesis \cite{BB-PRL}.

\section{Kinetic mixing of Ordinary and Mirror Photons } 

In the context of $G_{\rm SM} \times G_{\rm SM}'$, the 
Lagrangian can contain the gauge invariant mixing term 
between the field-strength tensors of the gauge
factors $U(1)$ and $U(1)'$. 
After the electroweak symmetry breaking, this term 
gives rise to a kinetic mixing term between the 
the O- and M-photons: 
\be{FFpr} 
\cL = -\eps F^{\mu\nu} F'_{\mu\nu}
\end{equation}
This term cannot be suppressed by symmetry reasons,  
and generally the constant $\eps$ could be of order 1. 

Once such a term is introduced, the following situation emerges. 
One has to diagonalize first the kinetic 
terms of the ordinary photon field $A_\mu$ and the mirror one 
$A'_\mu$, and identify the physical 
photon as a their linear combination. 
Now, once the kinetic terms are 
brought to canonical form by diagonalization and scaling 
of the fields, $(A,A')\to (A_1,A_2)$,  
any orthonormal combination of states $A_1$ and $A_2$ 
becomes good to describe the physical basis.  
In particular, $A_2$ can be chosen as a "sterile" state 
which does not couple to O-particles but only to M-particles. 
Then, the orthogonal combination $A_1$ couples not only 
to O-particles, but also with M-particles with a small 
charge $\propto 2\eps$ -- in other words, mirror matter 
becomes "milicharged" with respect to the physical ordinary 
photon \cite{Holdom,FYV}.

In this way, the term (\ref{FFpr}) induces the process
$e^+e^-\to e^{\prime+}e^{\prime-}$ with an amplitude just 
$2\eps$ times the $s$-channel amplitude for $e^+e^-\to e^+e^-$. 
This could have striking experimental implications for 
positronium physics:   
ordinary positronium mixes to its mirror counterpart which   
effect could be manifested as an invisible decay mode   
of the orthopositronium. Perhaps this effect could 
important for the troubling mismatch problems in the 
orthopositronium physics 
\cite{Glashow86,Gninenko}. 
For the moment, the experimental limits on the 
orthopositronium decays lead to an upper limit 
$\eps < 3\times 10^{-7}$ or so. 

The stronger limit can be obtained from the cosmology.  
As we already remarked, the BBN constraints require that 
mirror sector should be colder than the the ordinary one, 
$T'/T < 0.5$ or so. 
On the other hand, the reaction $e^+e^-\to e^{\prime+}e^{\prime-}$,   
funneling energy from O-sector to M-sector, 
would heat the latter too much before the BBN epoch,   
unless $\eps < 3\times 10^{-8}$ \cite{Glashow87}. 
  
The search of the process $e^+e^- \to$ {\it invisible} 
could approach sensitivities down to few $\times 10^{-9}$. 
\cite{Sergei} This interesting experiment could test 
the proposal of ref. \cite{Feet} claiming that the signal 
for the dark matter detection by the DAMA/NaI group \cite{Rita} 
can be explained by elastic scattering of M-baryons with 
ordinary ones mediated by kinetic mixing (\ref{FFpr}), 
if $\eps \sim 4\times 10^{-9}$. 

The smallness of the kinetic mixing term (\ref{FFpr}) can  
be naturally explained by invoking the concept of grand unification. 
Obviously, the term (\ref{FFpr}) is forbidden 
in GUTs like $SU(5)\times SU(5)'$. 
which do not contain abelian factors. 
However, given that both $SU(5)$ and $SU(5)'$ symmetries are 
broken down to their $SU(3)\times SU(2)\times U(1)$ subgroups 
by the Higgs 24-plets $\Phi$ and $\Phi'$, it could emerge from 
the higher order effective operator 
\be{GGpr}
\cL= -\frac{\zeta}{M^2} (G^{\mu\nu} \Phi) (G'_{\mu\nu} \Phi') 
\end{equation}
where $G_{\mu\nu}$ and $G'_{\mu\nu}$ are field-strength 
tensors respectively of $SU(5)$ and $SU(5)'$, and $M$ is 
some cutoff scale which can be of the order of $M_{Pl}$ or so. 
After substituting VEVs of $\Phi$ and $\Phi'$ 
the operator (\ref{FFpr}) is induced  with 
$\eps \sim \zeta (\langle\Phi\rangle/M)^2$. 

In fact, the operator (\ref{GGpr}) can be effectively induced 
by loop-effects involving some heavy fermion or scalar 
fields in the mixed representations of $SU(5)\times SU(5)'$, 
with $\zeta \sim \al/3\pi\sim 10^{-3}$ being a loop-factor.   
Consider, for example, $SU(5)\times SU(5)'$ theory 
which apart of the standard O- and M-fermion multiplets 
includes also also the chiral fermions in mixed representations 
$F\sim (5,5)$ and $F'\sim (\bar5,\bar5)$. 
These would necessarily appear if $SU(5)\times SU(5)'$ is 
embbedded into e.g. $SU(10)$ group. 
They should have a large mass term $MFF'$, e.g. of the order of 
$SU(10)$ breaking scale to $SU(5)\times SU(5)'$. 
However, in general they could have coupling terms 
$\Phi FF' + \Phi'FF'$  with the GUT Higgses 
$\Phi\sim (24,1)$ and $\Phi'\sim (1,24)$. 

With respect $G_{\rm SM}\times G_{\rm SM}'$ subgroup, 
these multiplets split into fragments $F_{ij}$ with different 
hypercharges $(Y_i,Y_j')$ with respect to $U(1)$ and $U(1)'$ 
factors, and correspondingly with masses 
$\hat{M}_{ij} = M+Y_i\langle\Phi\rangle + Y_j'\langle\Phi'\rangle$. 
Therefore, the loops involving 
the fermions $F_{ij}$ would induce a contribution to 
the term (\ref{FFpr}) with  
$\eps \simeq (\al/3\pi) {\rm Tr}[YY'\ln(\hat{M}/\La)]$ 
where $\La$ is an ultraviolet cutoff scale and under trace 
the sum over all fragments $F_{ij}$ is understood. 
As far as these fragments emerge from the GUT 
multiplets, they necessarily obey that ${\rm Tr}(YY')=0$, 
and thus $\eps$ should be finite and cutoff independent. 
Thus, expanding the logarithm in terms of small parameters 
$\langle\Phi^{(\prime)}\rangle/M$, we finally obtain 
\be{BB-rad}
\eps \simeq \frac{\al\langle\Phi\rangle \langle\Phi'\rangle}
{3\pi M^2} {\rm Tr}[(YY')^2]   
\end{equation} 
exactly what we expected from the effective operator 
(\ref{GGpr}). Hence, the heavy mixed multiplets in fact 
do not decouple and induce the O- and M-photon kinetic 
mixing term proportional to the square of typical mass 
splittings in these multiplets ($\sim \langle\Phi\rangle^2$), 
analogously to the familiar situation for the photon to 
$Z$-boson mixing in the standard model. 
Hence, taking the GUT scale as 
$\langle\Phi\rangle\sim 10^{16}$ GeV and $M\sim M_{Pl}$ 
we see that the strength of kinetic mixing term (\ref{FFpr}), 
could vary vary from $\eps \sim 10^{-10}$ to $10^{-8}$. 
Certainly, the coupling (\ref{GGpr}) can be stronger 
suppressed or completely eliminated by some symmetry reasons.

\section{Other possible interactions between O-and M-particles } 

Here we briefly discuss, what other common 
interactions and forces could exist between 
the O- and M-particles, including matter fields and 
gauge fields.  

It is pretty possible that O-and M-particles have 
common forces mediated by the gauge bosons of some 
additional symmetry group $H$. 
In other words,  
one can consider a theory with a gauge group 
$G\times G'\times H$, where O-particles are 
in some representations of $H$, $L_a\sim r_a$, and 
correspondingly their antiparticles are in antirepresentations, 
$\tR_a \sim \bar{r}_a$. 
As for M-particles, vice versa, we take $L'_a\sim \bar{r}_a$, 
and so $\tR'_a \sim r_a$. Only such a prescription of $\cG$ 
pattern is compatible with the mirror parity (\ref{MLR}). 
In addition, in this 
case $H$ symmetry automatically becomes vector-like and 
so it would have no problems with axial anomalies even 
if the particle contents of O- and M-sectors separately are 
not anomaly-free with respect to $H$. 

Let us consider the following example. The horizontal 
flavour symmetry $SU(3)_H$ between the quark-lepton families 
seems to be very promising for understanding the fermion 
mass and mixing pattern \cite{su3,PLB83}. 
In addition, it can be   
useful for controlling the flavour-changing phenomena 
in the context of supersymmetry \cite{PLB98}.      
One can consider e.g. a GUT with $SU(5)\times SU(3)_H$ 
symmetry where L-fermions in (\ref{LL-SM}) are triplets of
$SU(3)_H$. So $SU(3)_H$ has a chiral character and 
it is not anomaly-free unless some extra states are introduced 
for the anomaly cancellation \cite{su3}.  

However, the concept of mirror sector makes the things 
easier. 
Consider e.g. $SU(5)\times SU(5)'\times SU(3)_H$ theory  
with $L$-fermions being triplets of $SU(3)_H$ 
and $L'$-fermions anti-triplets. 
Hence, in this case the $SU(3)_H$ anomalies of the ordinary  
particles are cancelled by their mirror partners.  
Another advantage is that in a supersymmetric theory 
the gauge D-terms of $SU(3)_H$ are perfectly cancelled between 
the two sectors and hence they do not give rise to dangerous 
flavour-changing phenomena \cite{PLB98}.  
Common gauge $B-L$ symmetry between the two sectors  
can also be plausible. 

The immediate implication of interactions mediated by 
common gauge or Higgs bosons would be the 
mixing of neutral O-bosons to their M-partners, mediated 
by horizontal gauge bosons. Namely, oscillations 
$\pi^0 \to \pi^{\prime 0}$ or $K^0 \to K^{\prime 0}$ 
become possible and perhaps even detectable if the 
horizontal ($B-L$) gauge symmetry breaking scale is not too high. 

The operators of dimension 9 operators $(1/M^5)(udd)(u'd'd')$ 
would lead to oscilation between ordinary and mirror neutrons, 
$n \to n'$. Surprisingly, the experimental limits on such 
oscillation are very weak as compared to that of 
neutron-antineutron oscillation ($\tau_{n\bar n}> 10^8$ s), 
and allow the oscillation period as small $\tau_{nn'} \sim 1$ s, 
much smaller then the neutron lifetime \cite{n-npr}. 
This can make $n - n'$ oscillation easily detectable at 
"table-top" experiments and it can also have far going 
astrophysical implications.  
Remarkably, $\tau_{nn'}$ is not restricted by the limits on 
the nucleon stability. 

The model with common Peccei-Quinn symmetry between the 
O- and M-sectors was considered in \cite{BGG}. 
In this situation many astrophysical and cosmological 
bounds on the axion can be eliminated.  
Most interesting consequences follow if the mirror parity is 
broken, $v' \gg v$. In this case axion properties dramatically 
change (namely, relation between the axion mass and decay scale 
is strongly altered) and for a particular range of parameters 
one could have an axion with $f_a \sim 10^6$ GeV and 
$m_a\sim 1$ MeV, with interesting implications for the 
energetics of the Gamma Ray Bursts and Supernovae \cite{Drago}.

\section{The expansion of the Universe and 
thermodynamics of the O- and M-sectors}

Let us assume, that after inflation ended, the 
O- and M-systems received different reheating temperatures,  
namely $T_R> T'_R$. This is certainly possible despite 
the fact that two sectors have identical Lagrangians, and 
can be naturally achieved in certain 
models of inflation \cite{KST,BDM,BV}.\footnote{For analogy, 
two harmonic oscillators with the same frequency 
(e.g. two springs with the same material and the same length) 
are not obliged to oscillate with the same amplitudes.} 
  
If the two systems were decoupled already
after reheating, at later times $t$ they will have different
temperatures $T(t)$ and $T'(t)$, and so 
different energy and entropy densities:
\be{rho}
\rho(t) = {\pi^{2}\over 30} g_\ast(T) T^{4}, ~~~
\rho'(t) = {\pi^{2}\over 30} g'_\ast(T') T^{\prime4} ~,
\end{equation}
\be{s}
s(t) = {2\pi^{2}\over 45} g_{s}(T) T^{3} , ~~~
s'(t) = {2\pi^{2}\over 45} g'_{s}(T') T^{\prime3} ~.
\end{equation}
The factors $g_{\ast}$, $g_{s}$ and $g'_{\ast}$, $g'_{s}$
accounting for the effective number of the degrees of freedom
in the two systems can in general be different from each other.
Let us assume that during the expansion of the Universe 
the two sectors evolve with separately conserved entropies. 
Then the ratio $x\equiv (s'/s)^{1/3}$ is time independent 
while the ratio of the temperatures
in the two sectors is simply given by:
\be{t-ratio}
\frac{T'(t)}{T(t)} = x \cdot
\left[\frac{g_{s}(T)}{g'_{s}(T')} \right] ^{1/3} ~.
\end{equation}

The Hubble expansion rate is determined by the total
energy density $\rhb=\rho+\rho'$, $H=\sqrt{(8\pi/3) G_N\rhb}$.
Therefore, at a given time $t$ in a radiation dominated epoch
we have
\be{Hubble}
H(t) = {1\over 2t} = 1.66 \sqrt{\bg(T)} \frac{T^2}{M_{Pl}} =
1.66 \sqrt{\bg'(T')} \frac{T^{\prime2}}{M_{Pl}} ~
\end{equation}
in terms of O- and M-temperatures $T(t)$ and $T'(t)$, where
\beqn{g-ast}
\bg(T) = g_\ast (T) (1 + x^4), ~~~~
\bg'(T') = g'_\ast (T')(1 + x^{-4} ) .
\eeqn

In particular, we have  $x= T'_0/T_0$,
where $T_0,T'_0$ are the present
temperatures of the relic photons in O- and M-sectors.
In fact, $x$ is the only free parameter in our model
and it is constrained by the BBN bounds.

The observed abundances of light elements are in
good agreement with the standard nucleosynthesis predictions.
At $T\sim 1$ MeV we have $g_\ast=10.75$
as it is saturated by photons $\gamma$, electrons $e$  
and three neutrino species $\nu_{e,\mu,\tau}$.
The contribution of mirror particles
($\gamma'$, $e'$ and $\nu'_{e,\mu,\tau}$)
would change it to $\bg =g_\ast (1 + x^4)$.
Deviations from $g_\ast=10.75$ are usually
parametrized in terms of the effective number
of extra neutrino species,
$\Delta g= \bar{g}_\ast -10.75=1.75\cdot \Delta N_\nu$.
Thus we have:
\be{BBN}
\Delta N_\nu = 6.14\cdot x^4 ~.
\end{equation}
This limit very weakly depends on $\DN$.
Namely, the conservative bound $\Delta N_\nu < 1$ 
implies $x < 0.64$. 
In view of the present observational situation,
confronting the WMAP results to the BBN analysis, 
the bound seems to be stronger. 
However, e.g. $x = 0.3$ implies a completely negligible 
contribution  $\Delta N_\nu = 0.05$. 

As far as $x^4\ll 1$, in a relativistic epoch
the Hubble expansion rate (\ref{Hubble}) is dominated
by the O-matter density and the presence of the M-sector
practically does not affect the standard cosmology
of the early ordinary Universe.
However, even if the two sectors have the
same microphysics, the cosmology of the early
mirror world can be very different from the
standard one as far as the crucial epochs like 
baryogenesis, nuclesosynthesis, etc. are concerned.
Any of these epochs is related to an instant when
the rate of the relevant particle process
$\Ga(T)$, which is generically a function
of the temperature, becomes equal to the Hubble
expansion rate $H(T)$.
Obviously, in the M-sector these events take place
earlier than in the O-sector,
and as a rule, the relevant processes in the former
freeze out at larger temperatures than in the latter.

In the matter domination epoch the situation becomes
different.
In particular, we know that ordinary baryons provide
only a small fraction of the present matter density,
whereas the observational data
indicate the presence of dark matter with about 5 times 
larger density.  
It is interesting to question whether the missing
matter density of the Universe could be due to
mirror baryons? In the next section we show that   
this could occur in a pretty natural manner.

It can also be shown that the BBN epoch in the mirror
world proceeds differently from the ordinary one,
and it predicts different abundancies of primordial 
elements \cite{BCV}. 
It is well known that primordial abundances of the light 
elements depend on the baryon to photon density ratio 
$\eta=n_B/n_\gamma$, and the observational data well 
agree with the WMAP result $\eta \simeq 6\times 10^{-10}$. 
As far as $T'\ll T$, the universe expansion rate at 
the ordinary BBN epoch ($T\sim 1$ MeV) is determined by the 
O-matter density itself, and thus for the ordinary observer 
$\olga$ it would be very difficult to detect the contribution 
of M-sector: the latter is equivalent to 
$\Delta N_\nu \approx 6.14 x^4$ and hence it is 
negligible for $x\ll 1$. 
As for nucleosynthesis epoch in M-sector,  
the contribution of O-world instead is dramatic:  
it is equivalent to $\Delta N'_\nu \approx 6.14 x^{-4} \gg 1$. 
Therefore, mirror observer $\maxim$ which measures the abundancwes 
of mirror light elements should immediately observe discrepancy 
between the universe expansion rate and the M-matter density 
at his BBN epoch ($T'\sim 1$ MeV) as far as the former is 
determined by O-matter density which is invisible for 
$\maxim$. 
The result for mirror $^4$He also depends on the mirror baryon 
to photon density ratio $\eta'=n'_B/n'_\gamma$. 
Recalling that $\eta' = (\beta/x^3)\eta$, we see that 
$\eta \gg \eta$ unless $\beta = n'_B/n_B \ll x^3$. 
However, if $\beta >1$,  
we expect that mirror helium mass fraction $Y'_4$ would be 
considerably larger than the observable $Y_4\simeq 0.24$. 
Namely, direct calculations show that 
for $x$ varying from 0.6 to 0.1, $Y'_4$ would
varie in the range $Y'_4 =0.5-0.8$. 
Therefore, if M-baryons constitute dark matter or at least 
its reasonable fraction, the M-world is dominantly helium 
world while the heavier elements can also present with 
significant abundances. 

The `helium' nature of the mirror universe should have 
a strong impact on the processes of the star formation and 
evolution in the mirror sector \cite{Paolo}.

\begin{figure}[ht]
\centerline{\epsfxsize=3.9in\epsfbox{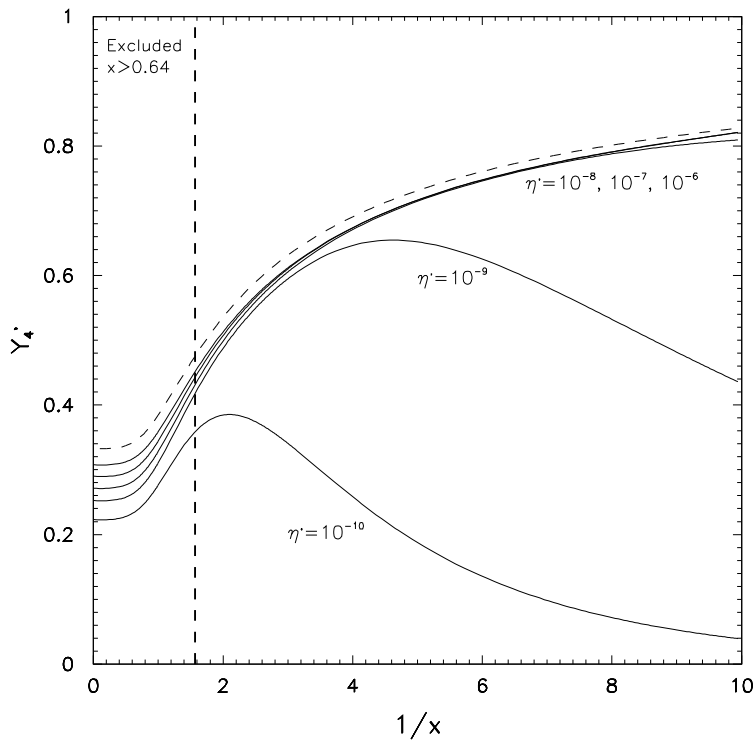}}
\caption{
The promordial mirror $^4$He mass fraction as a function of $x$. 
The different curves correspond to different fixed values of 
of $\eta'=n'_B/n'_\gamma$. 
The dashed curve extrapolates the case when mirror baryons 
constitute dark matter, i.e. $\beta\simeq 5$. 
\label{ian-bbn}}
\end{figure}

\section{Baryogenesis in M-sector and mirror baryons as 
dark matter} 

\subsection{Visible and dark matter in the Universe}

The present cosmological observations strongly support 
the main predictions of the inflationary scenario: 
first, the Universe is flat, with the energy density very close 
to the critical $\Om=1$, and second, 
primoridal density perturbations have nearly flat spectrum, 
with the spectral index $n_s\approx 1$. 
The non-relativistic matter gives only a small 
fraction of the present energy density, about 
$\Om_m \simeq 0.27$, while the rest is attributed to the 
vacuum energy (cosmological term) $\Om_\La \simeq 0.73$ 
\cite{WMAP}. 
The fact that $\Om_m$ and $\Om_\La$ are of the same order, 
gives rise to so called cosmological coincidence problem: 
why we live in an epoch when $\rho_m\sim\rho_\la$, if in the 
early Universe one had $\rho_m \gg \rho_\La$ and in the 
late Universe one would expect $\rho_m\ll \rho_\La$? 
The answer can be only related to an antrophic principle: 
the matter and vacuum energy densities 
scale differently with the expansion of the Universe 
$\rho_m\propto a^{-3}$ and $\rho_\La \propto$ const., 
hence they have to coincide at some moment, 
and we are just happy to be here. Moreover, for 
substantially larger $\rho_\La$ no galaxies could be formed 
and thus there would not be anyone to ask this this question. 

On the other hand, the matter in the Universe has two 
components, visible and dark: $\Om_m = \Om_b +\Om_{d}$.  
The visible matter consists of baryons  
with $\Om_b \simeq 0.044$ while the dark matter 
with $\Om_{d} \simeq 0.22$ is constituted by some hypothetic 
particle species very weakly interacting with the 
observable matter. 
It is a tantalizing question, why the visible and dark 
components have so close energy densities?  
Clearly, the ratio  
\be{b-dm}
\beta = \frac{\rho_d}{\rho_b}   
\end{equation}
does not depend on time as far as with the expansion of the 
Universe both $\rho_b$ and $\rho_{d}$ scale as $\propto a^{-3}$. 


In view of the standard cosmological paradigm,
there is no good reason for having $\Om_d \sim \Om_b$, 
as far as the visible and dark components have  
different origins. 
The density of the visible matter is $\rho_b =M_N n_b$,     
where $M_N\simeq 1$ GeV is the nucleon mass, and 
$n_b$ is the baryon number density of the Universe.   
The latter should be produced in a very early Universe 
by some baryogenesis mechanism, which is presumably related 
to some B and CP-violating physics at very high energies. 
The baryon number per photon $\eta=n_b/n_\ga$ is very small. 
Observational data on the primordial abundances of light 
elements and the WMAP results on the CMBR anisotropies nicely 
converge to the value $\eta \approx 6\times 10^{-10}$.  

As for dark matter, it is presumably constituted by 
some cold relics with mass $M$ and number density $n_{d}$, 
and $\rho_{d} = M n_{d}$. 
The most popular candidate for cold dark matter (CDM) 
is provided by the lightest supersymmetric particle (LSP) 
with $M_{\rm LSP}\sim 1$ TeV, 
and its number density $n_{\rm LSP}$ 
is fixed by its annihilation cross-section. 
Hence $\rho_b\sim \rho_{\rm LSP}$ requires that   
$n_b/n_{\rm LSP}\sim M_{\rm LSP}/M_N$ and the 
origin of such a conspiracy between four principally independent 
parameters is absolutely unclear.  
Once again, the value $M_N$ is fixed by the QCD scale while 
$M_{\rm LSP}$ is related to the supersymmetry breaking scale, 
$n_b$ is determined by B and CP violating properties of 
the particle theory at very high energies whereas  
$n_{\rm LSP}$ strongly depends on the supersymmetry 
breaking details. Within the parameter space of the MSSM 
it could vary within several orders of magnitude, and moreover, 
in either case it has nothing to do with the B and CP 
violating effects.  

The situation looks even more obscure if the dark component 
is related e.g. to the primordial oscillations 
of a classic axion field, in which case the dark matter 
particles constituted by axions are superlight, with mass 
$\ll 1$ eV, but they have a condensate with enormously 
high number density. 

In this view, the concept of mirror world could give 
a new twist to this problem. Once the visible matter 
is built up by ordinary baryons, then the mirror 
baryons could constitute dark matter in a natural way.  
They interact with mirror photons, however they are dark 
in terms of the ordinary photons. 
The mass of M-baryons is the same as the ordinary 
one, $M=M_N$, and so we have $\beta = n'_b/n_b$, 
where $n'_b$ is the number density of M-baryons. 
In addition, as far as the two sectors have the same 
particle physics, it is natural to think that the M-baryon 
number density $n'_b$ is determined by the baryogenesis 
mechanism which is similar to the one which fixes the 
O-baryon density $n_b$. Thus, one could question whether 
the ratio $\beta=n'_b/n_b$ could be naturally order 1 
or somewhat bigger.     

The visible matter in the Universe consists of baryons, 
while the abundance of antibaryons is vanishingly small. 
In the early Universe, at tempreatures $T\gg 1$ GeV, 
the baryons and antibaryons had practically the same 
densities, $n_b\approx n_{\bar b}$ with $n_b$ slightly 
exceeding $n_{\bar b}$, so that 
the baryon number density was small, 
$n_B=n_b-n_{\bar b}\ll n_b$. 
If there was no significant entropy production after 
the baryogenesis epoch,  
the baryon number density to entropy density 
ratio had to be the same as today,  
$B = n_B/s \approx 8\times 10^{-11}$.\footnote{
In the following we use $B=n_B/s$ which is related  
with the familiar $\eta=n_B/n_\ga$ as $B\approx 0.14\eta$. 
However, $B$ is more adopted for featuring the baryon 
asymmetry since it does not depend on time if the entropy of 
the Universe is conserved.}
 
One can question, who and how has prepared the initial 
Universe with such a small excess of baryons over antibaryons. 
In the Friedman Universe the initial baryon asymmetry 
could be arranged a priori, in terms of non-vanishing 
chemical potential of baryons. 
However, the inflationary paradigm gives another 
twist to this question, since inflation dilutes any 
preexisting baryon number of the Universe to zero.  
Therefore, after inflaton decay and the (re-)heating of the Universe, 
the baryon asymmetry has to be created by some cosmological 
mechanism.  

There are several relatively honest baryogenesis mechanisms 
as are GUT baryogenesis, leptogenesis, electroweak baryogenesis, 
etc. (for a review, see e.g. \cite{BA-Dolgov}). 
They are all based on general principles 
suggested long time ago by Sakharov \cite{Sakh}:  
a non-zero baryon asymmetry can be produced in the initially baryon
symmetric Universe if three conditions are fulfilled: B-violation,
C- and CP-violation and departure from thermal equilibrium.
In the GUT baryogenesis or leptogenesis scenarios 
these conditions can be satisfied in the decays of heavy 
particles. 

At present it is not possible to say definitely which 
of the known mechanisms is responsible 
for the observed baryon asymmetry in the ordinary world. 
However, it is most likely that the baryon asymmetry in 
the mirror world 
is produced by the same mechanism
and moreover, the properties of the $B$ and CP violation
processes are parametrically the same in both cases.
But the mirror sector has a lower temperature than ordinary one, 
and so at epochs relevant for baryogenesis the out-of-equilibrium 
conditions should be easier fulfilled for the M-sector.


\subsection{Baryogenesis in the O- and M-worlds} 

Let us consider the difference between the ordinary and 
mirror baryon asymmetries on the example of the 
GUT baryogenesis mechanism. It is typically based on
`slow' B- and CP-violating decays of a superheavy boson $X$
into quarks and leptons, where slow means that    
at $T < M$ the Hubble parameter $H(T)$ is greater than   
the decay rate $\Gamma \sim \alpha M$, 
$\alpha$ being the coupling strength of $X$ to fermions
and $M$ its mass. 
The other reaction rates are also of relevance: 
{\it inverse decay}: $\Gamma_{I} \sim  \Gamma    
(M/T)^{3/2} \exp(-M/T)$ for $T< M_X$, 
and {\it the $X$ boson mediated scattering processes}:
$\Gamma_S \sim n_X \sigma\sim  A \alpha^2T^5/M^4$,
where the factor $A$ amounts for the possible reaction channels.

The final BA depends on the temperature
at which $X$ bosons go out from equilibrium.
One can introduce a parameter which measures the
effectiveness of the decay at the
epoch $T\sim M$:
$k=(\Gamma/H)_{T=M}= 0.3\bg^{-1/2}(\alpha M_{Pl}/M)$.
For $k \ll 1$ the out-of-equilibrium condition is 
strongly satisfied, and per decay of one $X$ particle 
one generates the baryon number proportional to the 
CP-violating asymmetry $\eps$. Thus,  
we have $B = \eps/g_\ast$, 
$g_\ast$ is a number of effective degrees of freedom at $T < M$.   
The larger $k$ is, the longer equilibrium is
maintained and the freeze-out abundance of $X$ boson
becomes smaller. 
Hence, the resulting
baryon number to entropy ratio becomes  
$B=(\eps/g_\ast)D(k)$, where the damping factor $D(k)$  
is a decreasing function of $k$. 
In particular, $D(k)=1$ for $k\ll 1$, while 
for $k$ exceeding some critical value 
$k_c$, the damping is exponential.  

The presence of the mirror sector practically  
does not alter the ordinary baryogenesis.
The effective particle number is
$\bar g_\ast (T) = g_\ast(T)(1+x^4)$ and thus
the contribution of M-particles to the Hubble constant  
at $T\sim M$ is suppressed by a small factor $x^4$.

In the mirror sector everything should occur
in a similar way, apart from the fact that now  
at $T'\sim M$ the Hubble constant is not
dominated by the mirror species but by ordinary ones:
$\bar g'_\ast (T')\simeq g'_\ast (T')(1+ x^{-4})$.
As a consequence, we have
$k' = (\Gamma/H)_{|T'=M} = k x^2$.
Therefore, the damping factor for mirror baryon asymmetry 
can be simply obtained by replacing 
$k\rightarrow k'=kx^2$ in $D(k)$. 
In other words, the baryon number density to entropy density 
ratio in the M-world becomes 
$B'=n'_B/s' \simeq (\epsilon/g_\ast) D(kx^2)$.
Since $D(k)$ is a decreasing function of $k$, then
for $x < 1$ we have $D(kx^2) > D(k)$ and
thus we conclude that the mirror world always gets a
{\it larger} baryon asymmetry than the visible one, 
$B' > B$.\footnote{As it was shown in ref. \cite{BCV}, 
the relation $B' > B$ takes place also in the context 
of the electroweak baryogenesis scenario, where the 
out-of-equilibrium conditions is provided by fast 
phase transition and bubble nucleation.} 
Namely, for $k>1$ the baryon asymmetry in the O-sector is 
damped by some factor -- we have 
$B\simeq (\eps/g_\ast)D(k) < \eps/g_\ast$, 
while if $x^2 < k^{-1}$, the damping would be irrelevant 
for the M-sector and hence $B'\simeq \eps/g_\ast$.  

However, this does not a priori mean
that $\Omega'_b$ will be larger than $\Omega_b$.
Since the entropy densities are related as $s'/s=x^3$,
for the ratio $\beta =\Omega'_b/\Omega_b$ we have:
\be{B-ratio}
\beta(x) =  \frac{n'_B}{n_B} = \frac{B's'}{Bs} =
\;\frac{x^3D(kx^2)}{D(k)} ~.
\end{equation}
The behaviour of this ratio 
as a function of $k$ for different values of the
parameter $x$ is given in the ref. \cite{BCV}.
Clearly, in order to have $\Omega'_b > \Omega_b$, 
the function $D(k)$ has to decrease
faster than $k^{-3/2}$ between $k'=kx^2 $ and $k$.
Closer inspection of this function reveals
that the M-baryons can be overproduced only if
$k$ is sufficiently large, so that 
the relevant interactions in the observable sector
maintain equilibrium longer than in the mirror one,
and thus ordinary BA can be suppressed by an
exponential Boltzmann factor while the mirror BA  
could be produced still in the regime $k' =kx^2 \ll 1$, 
when $D(k')\approx 1$. 

However, the GUT baryogenesis picture has the 
following generic problem. 
In scenarios based on grand unification models like $SU(5)$, 
the heavy gauge or Higgs boson decays violate separately 
$B$ and $L$, but conserve $B-L$, and so finally $B-L=0$.
On the other hand, the non-perturbative sphaleron processes, 
which violate $B+L$ but conserve $B-L$,
are effective at temperatures from about $10^{12}$ GeV down to 
100 GeV \cite{KRS}. 
Therefore, if $B+L$ is erased by sphaleron transitions, 
the final $B$ and $L$ both will vanish. 

Hence, in a realistic scenario one actually has to produce 
a non-zero $B-L$ rather than just a non-zero $B$, 
a fact that strongly favours the so called {\sl leptogenesis} 
scenario \cite{FY}.
The seesaw mechanism for neutrino masses offers an elegant   
possibility of generating non-zero $B-L$ in CP-violating decays  
of heavy Majorana neutrinos $N$ into leptons and Higgses. 
These decays violate $L$ but obviously do not change $B$  
and so they could create a non-zero $B-L = - L_{\rm in}$.   
Namely, due to complex Yukawa constants, the decay rates
$\Gamma(N\to l\phi)$ and $\Gamma(N \to \tl\tphi)$      
can be different  from each other, so that the leptons $l$ and
anti-leptons $\tl$ are produced in different amounts.

When sphalerons are in equilibrium, they violate $B+L$ and 
so redistribute non-zero $B-L$ between the 
baryon and lepton numbers of the Universe. Namely, the final values  
of $B$ and $B-L$ are related as $B = a(B-L)$, where 
$a$ is order 1 coefficient, namely $a\simeq 1/3$ in the 
SM and in its supersymmetric extension \cite{BA-Dolgov}. 
Hence, the observed baryon to entropy density ratio,
$ B \approx 8 \times 10^{-11} $,
needs to produce $ B-L \sim 2\times 10^{-10} $.

However, the comparative analysis presented above for the 
GUT baryogenesis in the O- and M-worlds, is essentially
true also for the leptogenesis scenario. 
The out-of-equilibrium decays of heavy $N$ neutrinos of 
the O-sector would produce a non-zero $B-L$ which being reprocessed 
by sphalerons would give an observable baryon asymmetry 
$B=a(B-L)$. On the other hand, the same decays of heavy 
$N'$ neutrinos of the M-sector would give non-zero $(B'-L')$  
and thus the mirror baryon asymmetry $B'=a(B'-L')$. 
In order to thermally produce heavy neutrinos in 
both O- and M-sectors, the lightest of them should 
have a mass smaller than the reheating temperature $T'_R$ 
in the M-sector, i.e. $M_N < T'_R ,T_R$. 
The situation $M_N > T'_R$ 
would prevent thermal production of $N'$ states, 
and so no $B'-L'$ would be generated in M-sector.  
However, one can 
consider also scenarios when both $N$ and $N'$ states 
are non-thermally produced in inflaton decays, but 
with different amounts. Then the reheating of both sectors 
as well as $B-L$ number generation can be related to the 
decays of the heavy neutrinos of both sectors and hence 
the situation $T'_R < T_R$ can be naturally accompanied by 
$B' > B$.

\section{
Baryogenesis via Ordinary-Mirror particle interaction } 

\begin{flushleft}
{\sl  Tweedledum and Tweedledee \\
Agreed to have a battle; \\
For Tweedledum said Tweedledee \\
Had spoiled his nice new rattle.} 
\end{flushleft}

An alternative mechanism of leptogenesis is based on scattering
processes rather than on decay \cite{BB-PRL}.
The main idea consists in the following.
The hidden (mirror) sector of particles is not in 
thermal equilibrium with the ordinary particle
world as far as the two systems interact very weakly. 
However, superheavy singlet neutrinos can mediate very weak 
effective interactions between the ordinary and mirror leptons.
Then, a net $B-L$ could emerge in the Universe as
a result of CP-violating effects in the unbalanced production
of mirror particles from ordinary particle collisions.     

As far as the leptogenesis is concerned, we concentrate only
on the lepton sector of both O and M worlds.
Therefore we consider the standard model, and among
other particles species, concentrate on 
the lepton doublets $l_i = (\nu, e)_i$
($ i=1,2,3 $ is the family index)
and the Higgs doublet $ \phi $ for the O-sector, 
and on their mirror partners 
$ l'_i = (\nu', e')_i $ and $ \phi' $. 
Their couplings to the heavy singlet neutrinos are 
given by (\ref{Yuk-nu}). 

Let us discuss now in more details this mechanism. 
A crucial role in our considerations is played by the reheating
temperature $T_R$, at which the inflaton decay and entropy production
of the Universe is over, and after which the Universe is dominated by
a relativistic plasma of ordinary particle species.
As we discussed above, we assume that after the postinflationary
reheating, different temperatures are established in the two
sectors: $T'_R < T_R$,  i.e. the mirror sector is cooler than
the visible one, or ultimately, even completely 
``empty".\footnote{It should be specified that $T_R,T'_R$ 
mean the temperatures at the moment when the energy density 
of relativistic products of the inflaton decay started to dominate 
over the energy density of the inflaton oscillation.}  

In addition, the two particle systems should interact very weakly
so that they do not come in thermal equilibrium with each other
after reheating.
We assume that the heavy neutrino masses are larger than 
$T_R$ and thus cannot be thermally produced.
As a result, the usual leptogenesis mechanism via
$N\to l\phi$ decays is ineffective.

Now, the important role is played by lepton number violating
scatterings mediated by the heavy neutrinos $ N $.
The ``cooler" mirror world starts to be ``slowly" occupied
due to the entropy transfer from the ordinary sector through   
the $ \Delta L=1 $ reactions
$ l_i \phi \to \bar \lpr_k \barphpr $,
$ \bar l_i \barphi \to \lpr_k \phpr $.
In general these processes violate CP due to complex
Yukawa couplings in eq.~(\ref{Yuk-nu}), and so the
cross-sections with leptons and anti-leptons in the initial state
are different from each other.
As a result, leptons leak to the mirror sector with different 
rate than antileptons and so a non-zero $B-L$ is produced
in the Universe. 

It is important to stress that this mechanism would generate
the baryon asymmetry not only in the observable sector,
but also in the mirror sector. In fact, the two sectors are completely
similar, and have similar CP-violating properties.
We have scattering processes which transform the ordinary particles
into their mirror partners, and CP-violation effects in this
scattering owing to the complex coupling constants.
These exchange processes are active at some early epoch of the 
Universe, and they are out of equilibrium.
In this case, at the relevant epoch, 
ordinary observer $\olga$ should detect that  
(i) matter slowly (in comparison to the Universe expansion rate)  
disappears from the thermal bath of O-world,  
(ii) particles and antiparticles disappear with different rates,
and at the end of the day she observes that her world
acquired a non-zero baryon number even if initially it was 
baryon symmetric.

On the other hand, his mirror colleague $\maxim$  would see that
(i) matter creation takes place in M-world, 
 (ii) particles and antiparticles appear with different rates.  
Therefore,  he also would observe that 
a non-zero baryon number is induced in his world.

One would naively expect that in this case the baryon asymmetries
in the O- and M-sectors should be literally equal,
given that the CP-violating factors are the same for both sectors.
However, we show that in reality, the BA in the M sector,
since it is colder, can be about an order of magnitude bigger   
than in the O sector, as far as washing out effects are taken into account.
Indeed, this effects should be more efficient for the hotter O-sector
while they can be negligible for the colder M sector, and 
this could provide reasonable differences between the two worlds
in case the exchange process is not too far from equilibrium.
The possible marriage between dark matter and the leptobaryogenesis
mechanism is certainly an attractive feature of our scheme.

The reactions relevant for the O-sector are the $\Delta L=1$
one $ l\phi \to \bar \lpr\barphpr $, and the $\Delta L=2$
ones like $l\phi \to \barl\barphi$, $ll\to \phi\phi$ etc.
Their total rates are correspondingly 
\beqn{rates} 
\Ga_{1} = \frac{Q_{1}n_{\rm eq} }{8\pi M^2};  ~~~~ 
Q_1={\rm Tr}(D^\dagger D) =
{\rm Tr}[(y^{\prime\dagger}y^\prime)^\ast g^{-1}
(y^\dagger y) g^{-1} ] ,  \nonumber \\
\Ga_{2} = \frac{3Q_{1}n_{\rm eq}}{4\pi M^2};  ~~~~ 
Q_2={\rm Tr}(A^\dagger A) =
{\rm Tr}[(y^{\dagger}y)^\ast g^{-1}(y^\dagger y) g^{-1}] , 
\eeqn
where $n_{\rm eq}\simeq (1.2/\pi^2)T^3$ is an equilibrium density  
per one bosonic degree of freedom, and the sum is taken over 
all isospin and flavour indices of initial and final states.  
It is essential that these processes stay out of equilibrium,  
which means that their rates should not exceed much 
the Hubble parameter $H = 1.66 \, g_\ast^{1/2} T^2 / M_{Pl} $
for temperatures $T \leq T_R$, where 
$g_\ast$ is the effective number of particle degrees of freedom, 
namely $g_\ast \simeq 100$ in the SM. 
In other words, the dimensionless parameters 
\beqn{K12} 
k_1 = \left({{\Gamma_1} \over {H}}\right)_{T=T_R} 
\simeq 1.5 \times 10^{-3}\, {{Q_1 T_R M_{Pl}} \over 
{g_\ast^{1/2}M^2}}  \,  \nonumber \\
k_2 = \left({{\Gamma_2} \over {H}}\right)_{T=T_R} 
\simeq 9 \times 10^{-3}\, {{Q_2 T_R M_{Pl}} \over 
{g_\ast^{1/2}M^2}}  \, 
\eeqn
should not be much larger than 1.

\begin{figure}[t]
  \begin{center}
    \leavevmode
    \epsfxsize = 7cm
    \epsffile{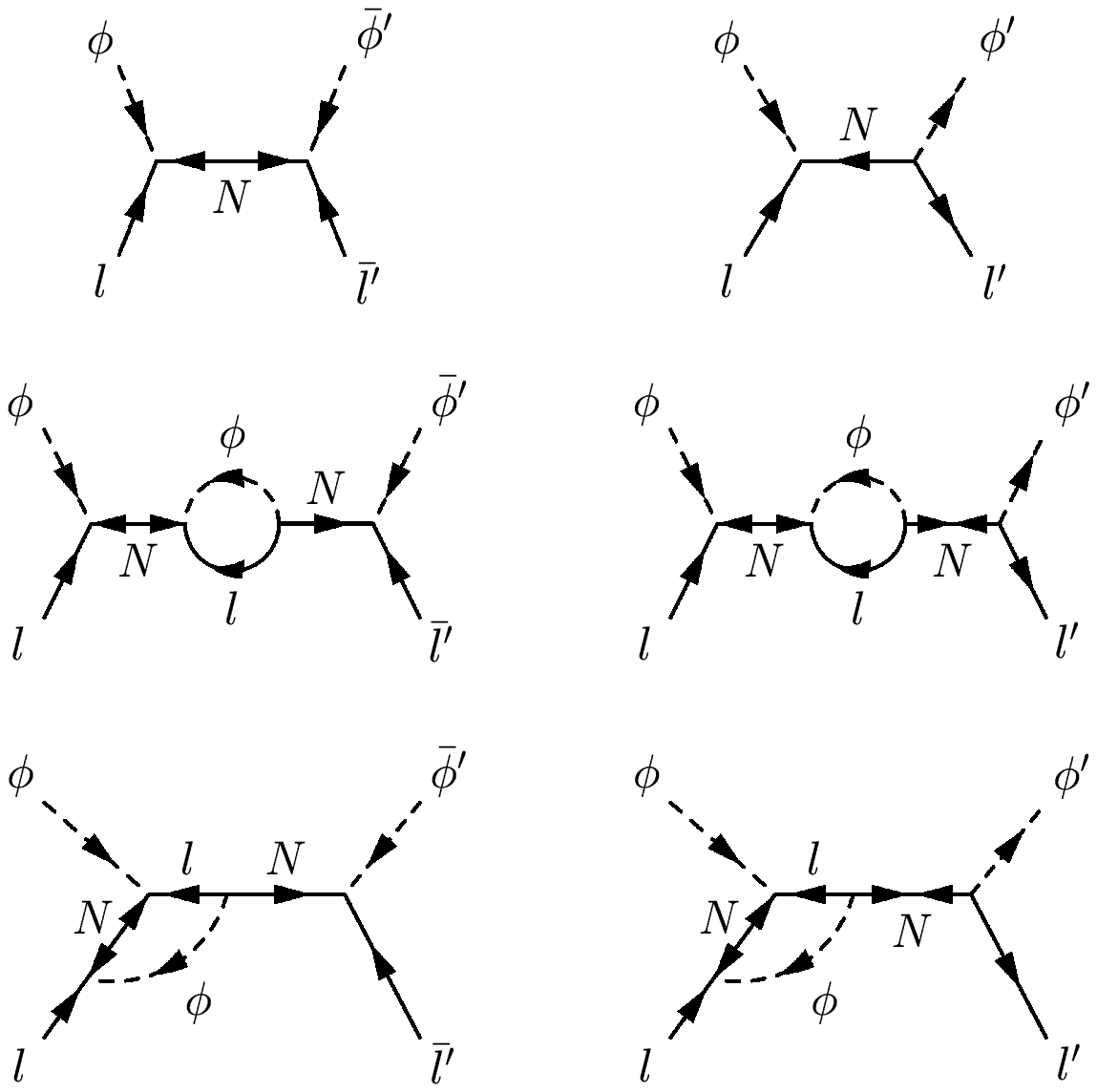}
  \end{center}
\caption{\small Tree-level and one-loop
diagrams contributing to the CP-asymmetries in
$l \phi \to \barlpr \barphpr$ (left column) and  
$l \phi \to \lpr \phpr$ (right column).}
\label{fig1}
\end{figure}

\begin{figure}[t]
  \begin{center}
    \leavevmode
    \epsfxsize = 7cm
    \epsffile{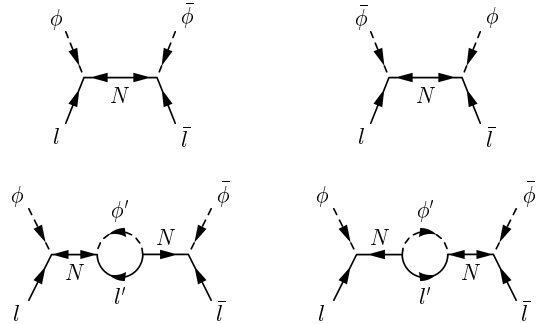}
  \end{center}
\caption{\small Tree-level and one-loop diagrams contributing  
to the CP-asymmetry of $l \phi \to \barl \barphi$.
The external leg labels identify the initial and final state particles.}
\label{fig2}
\end{figure}

Let us now turn to CP-violation.
In $\Delta L=1$ processes the CP-odd lepton number asymmetry
emerges from the interference between the tree-level and
one-loop diagrams of fig.~\ref{fig1}. 
However, CP-violation takes also place in $\Delta L=2$ processes 
(see fig.\ \ref{fig2}).
This is a consequence of the very existence of the mirror sector,
namely, it comes from the contribution of the mirror particles 
to the one-loop diagrams of fig.\ \ref{fig2}. 
The direct calculation gives:\footnote{
It is interesting to note that the tree-level amplitude for the 
dominant channel $l\phi\to \barlpr\barphpr$ goes as $ 1/M$ and the
radiative corrections as $ 1/M^3$.
For the channel $l\phi\to \lpr\phpr$ instead, both tree-level and
one-loop amplitudes go as $ 1/M^2$.
As a result,  the cross section CP asymmetries are comparable 
for both $l\phi\to \barlpr\barphpr$ and
$l\phi\to \lpr\phpr$ channels. } 
\begin{eqnarray}\label{CP}
&&
\sigma (l\phi\to \barl\barphi) -   
\sigma(\barl\barphi \to l\phi) = \Delta\sigma \, ;  \nonumber \\
&&
\sigma (l\phi\to \barlpr\barphpr) -
\sigma(\barl\barphi \to \lpr\phpr) =
(- \Delta\sigma  - \Delta\sigma' ) /2
\, ,  \nonumber \\
&&
\sigma (l\phi\to \lpr\phpr) -
\sigma(\barl\barphi \to \barlpr\barphpr) =
( -\Delta\sigma + \Delta\sigma' )/2
\, ,   \nonumber  \\
&&
\Delta\sigma = {{3J\, S} \over {32\pi^2 M^4}} \, , ~~~~~
\Delta\sigma' = {{3J'\, S} \over {32\pi^2 M^4}} \, ,
\end{eqnarray}
where $S$ is the c.m.\ energy square,  
\be{J}
J= {\rm Im\, Tr} [ (y^{\dagger}y)^\ast g^{-1}
(y^{\prime\dagger}y^\prime) g^{-2} (y^\dagger y) g^{-1}]  
\end{equation}
is the CP-violation parameter and
$J'$ is obtained from $J$ by exchanging
$y$ with $y^\prime$. The contributions yielding asymmetries 
$\mp \Delta \sigma'$ respectively for $l\phi\to \barlpr\barphpr$ and
$l\phi\to \lpr\phpr$ channels emerge from the diagrams with $\lpr\phpr$ 
inside the loops, not shown in fig.\ \ref{fig1}.

Of course, this is in agreement with CPT theorem that requires
that the total cross sections for particle and anti-particle
scatterings are equal to each other:
$\sigma(l\phi \to X) = \sigma(\barl\barphi \to X)$.
Indeed, taking into account that
$\sigma(l\phi \to l\phi) = \sigma(\barl\barphi \to \barl\barphi)$
by CPT, we see that CP asymmetries in the $\Delta L=1$ and
$\Delta L=2$ processes should be related as
\beqn{CPT}
&& 
\sigma (l\phi\to \barl\barphi) - \sigma(\barl\barphi \to l\phi) = 
\Delta \sigma   \;, \nonumber \\ 
&& 
\sigma(l\phi \to X^\prime) - \sigma(\barl\barphi \to X^\prime) =
- \Delta \sigma   \;, 
\eeqn
where $X^\prime$ are the mirror sector final states, either 
$\barlpr\barphpr$ or $\lpr\phpr$.
That is, the  $\Delta L=1$ and $\Delta L=2$ reactions have
CP asymmetries with equal intensities but opposite signs.

But, as $L$ varies in each case by a different amount, a net
lepton number decrease is produced, or better, a net increase
of $B-L$ $ \propto \Delta\sigma$
(recall that the lepton number $L$ is violated by the sphaleron
processes, while $B-L$ is changed 
solely by the above processes).

As far as we assume that the mirror sector is cooler and thus
depleted of particles, the only relevant reactions are the ones with
ordinary particles in the initial state. Hence,  the evolution of
the $B-L$ number density is determined by the CP asymmetries   
shown in eqs.~(\ref{CP}) and obeys the equation
\be{L-eq-2}
{{ d n_{B-L} } \over {dt}} + 3H n_{B-L} + \Gamma n_{B-L} =
\frac34 \Delta\sigma \, n_{\rm eq}^2 = 
1.8 \times 10^{-3}\, \frac{T^8}{M^4}  \;,
\end{equation}
where $ \Gamma = \Gamma_1 + \Gamma_2$ 
is the total rate of the $\Delta L=1$ and $\Delta L=2$ reactions, 
and for the CP asymmetric cross section $\Delta\sigma$ we take
the thermal average c.m.\ energy square $S \simeq 17\, T^2$. 

It is instructive to first solve this equation in the limit 
$k_{1,2} \ll 1$, when the out-of-equilibrium conditions 
are strongly satisfied and thus the term $\Gamma n_{B-L}$ 
can be neglected. Integrating this equation we obtain 
for the final $B-L$ asymmetry of the Universe, 
$Y_{BL} = n_{B-L}/s$, 
where $s=(2\pi^2/45)g_\ast T^3$ is the entropy density, 
the following expression:\footnote{ 
Observe that the magnitude of the produced
$B-L$ strongly depends on the temperature, namely,
larger $B-L$ should be produced in the patches where the plasma
is hotter. 
In the cosmological context, this would lead to a situation
where, apart from the adiabatic density/temperature perturbations,
there also emerge correlated isocurvature fluctuations with variable
$B$ and $L$ which could be tested with the future data on the
CMB anisotropies and large scale structure.}
\be{BL}
Y_{BL}^{(0)} \approx 2 \times 10^{-3} \,
{{J\, M_{Pl} T_R^3} \over {g_\ast^{3/2} M^4 }} . 
\end{equation}
It is interesting to note that 3/5 of this value 
is accumulated at temperatures $T>T_R$ 
and it corresponds to the amount of $B-L$ produced 
when the inflaton field started to decay and the
particle thermal bath was produced 
(Recall that the maximal temperature at the reheating period is
usually larger than $T_R$.)
In this epoch the Universe was still dominated by the inflaton
oscillations and therefore it expanded as $a\propto t^{2/3}$ 
while the entropy of the Universe was growing as $t^{5/4}$.
The other 2/5 of (\ref{BL}) is produced at $T<T_R$, 
in radiation dominated era when the Universe expanded as 
$a\propto t^{1/2}$ with conserved entropy (neglecting 
the small entropy leaking from the O- to the M-sector).   

This result (\ref{BL}) can be recasted as follows
\be{B-L}
Y_{BL}^{(0)} \approx 
\frac{20 J k^2 T_R}{g_\ast^{1/2} Q^2 M_{Pl}}    
\approx 10^{-10} \, \frac{Jk^2 }{Q^2} 
\left(\frac{T_R}{10^9~ {\rm GeV}}\right)   
\end{equation}
where $Q^2=Q_1^2+Q_2^2$, $k=k_1+k_2$ and 
we have taken again $g_\ast \approx 100$.
This shows that for Yukawa constants spread e.g.\ in the range
$ 0.1-1 $, one can achieve $ B-L = {\cal O}(10^{-10}) $
for a reheating temperature as low as $ T_R\sim 10^9 $ GeV.
Interestingly, this coincidence with the upper bound from the
thermal gravitino production, $ T_R < 4\times 10^9 $ GeV or so
\cite{Ellis}, indicates that our scenario could also work
in the context of supersymmetric theories.

Let us solve now eq. (\ref{L-eq-2}) exactly, without assuming 
$\Ga \ll H$. In this case we obtain \cite{BBC}: 
\be{BLO} 
Y_{BL}  =  D(k) \cdot Y_{BL}^{(0)} \;,
\end{equation}
where 
the depletion factor $D(k)$ is given by
\be{Dk}
D(k) = \frac35\, e^{-k} F(k) + \frac25\, G(k) 
\end{equation}
where 
\beqn{Fy}
&& 
F(k) = \frac{1}{4 k^4} \left[(2k -1)^3 + 6k-5 +6 e^{-2k}\right] , 
\nonumber \\
&& 
G(k) = \frac{3}{k^3} \left[ 2 -(k^2 + 2k +2) e^{-k} \right] . 
\eeqn
These two terms in $D(k)$ 
correspond to the integration of (\ref{L-eq-2}) respectively
in the epochs before and after reheating 
($T > T_R$ and $T < T_R$).
Obviously, for $ k \ll 1 $ we have 
$ D(k) = 1 $ and thus we recover the result as in (\ref{BL}) or
(\ref{B-L}): 
$
 = (B-L)_0 $.
However, for large $k$ the depletion can be reasonable (see. Fig. 5), 
e.g. for $k=1,2$ we have respectively $D(k) = 0.35, 0.15$.

Now, let us discuss how the mechanism considered above produces
also the baryon prime asymmetry in the mirror sector.
The amount of this asymmetry will depend on the CP-violation
parameter $ J^\prime = {\rm Im\, Tr}
[ (y^\dagger y) g^{-2} (y^{\prime\dagger} y^\prime)
g^{-1} (y^{\prime\dagger} y^\prime)^\ast g^{-1}] $ 
that replaces $J$ in $ \Delta \sigma' $ of eqs.\ (\ref{CP}). 
The mirror P parity under the exchange $ \phi \to \phi^{\prime\dagger} $,
$ l \to \bar{l}^\prime $, etc.,
implies that the Yukawa couplings are essentially the same in
both sectors, $ y^\prime= y^\ast$.
Therefore, in this case also the CP-violation parameters are
the same, $ J^\prime = -J $.\footnote{It is interesting to remark 
that this mechanism needs the left-right parity $P$ rather than the 
direct doubling  one $D$. One can easily see that the 
latter requires $y^\prime= y$, and so the CP-violating 
parameters $J$ and $J'$ are both vanishing.}  
Therefore, one naively expects that
$ n'_{B-L} = n_{B-L} $ and the mirror baryon density 
should be equal to the ordinary one,
$ \Omega'_{b} = \Omega_{b} $.

However, now we show that if the $ \Delta L = 1 $ and
$ \Delta L = 2 $ processes are not very far from equilibrium,
i.e. $ k \sim 1 $, the mirror baryon density should
be bigger than the ordinary one.
Indeed, the evolution of the mirror $B-L$  number density, 
$n'_{B-L}$, obeys the equation
\be{L-eq-3}
{{ d n'_{B-L} } \over {dt}} + 3H n'_{B-L} + \Gamma' n'_{\rm B-L} =
  {3 \over 4} \Delta\sigma' \, n_{\rm eq}^2   \; ,
\end{equation}
where now $ \Gamma' = ( Q_1 + 6Q_2 ) n'_{\rm eq} / 8\pi M^2 $
is the total reaction rate of the
$ \Delta L' = 1 $ and $ \Delta L' = 2 $ processes in the mirror sector,
and $ n'_{\rm eq} = (1.2 / \pi^2) T^{\prime 3} = x^3 n_{\rm eq} $
is the equilibrium number density per degree of freedom in the 
mirror sector. Therefore $k'=\Gamma'/H = x^3 k $,
and for the mirror sector we have
$ Y'_{BL} = D(kx^3)\cdot Y_{BL}^{(0)}$.  
Hence,  if $kx^3 \ll 1$,   
the depletion can be irrelevant: $D(kx^3)\approx 1$. 

Now taking into the account that in both sectors the
$B-L$ densities are reprocessed into the baryon number densities
by the same sphaleron processes,
we have $B = a(B-L)$ and $B' = a(B-L)'$,
with coefficients $a$ equal for both sectors.
Therefore, we see that the cosmological densities of the ordinary 
and mirror baryons should be related as
\be{omegabp}
\beta = \frac{\Om'_b}{\Om_b} \approx \frac{1}{D(k)}
\end{equation}
If $k \ll 1$,
depletion factors in both sectors are $ D \approx D' \approx 1 $
and thus we have that the mirror and ordinary baryons have the
same densities,
$\Omega'_{\rm b} \approx \Omega_{\rm b}$.
In this case mirror baryons are not enough to explain all dark 
matter and one has to invoke also some other kind of dark matter,
presumably cold dark matter.

However, if $k \sim 1$, then we would have
$ \Omega'_{\rm b} > \Omega_{\rm b} $,
and thus all dark matter of the Universe could be in the form
of mirror baryons.
Namely, for $k \simeq1.5$ we would have from eq.~(\ref{omegabp})
that $\Omega'_{b}/\Omega_{b} \approx 5$,
which is about the best fit relation between the ordinary 
and dark matter densities.

\begin{figure}[ht]
\centerline{\epsfxsize=3.9in\epsfbox{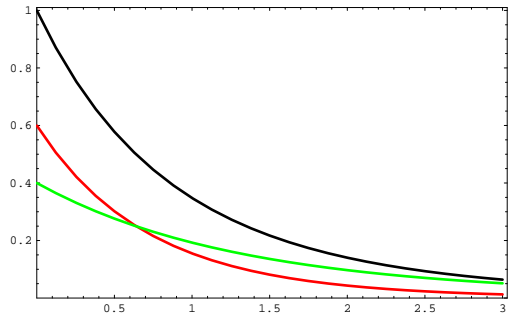}}
\vspace{-80mm}
\caption{
Damping factor $D(k)\approx\Omega_b/\Omega'_b$ (upper curve) and 
`constituent' functions $F(k)$ and $G(k)$ (lower curves). 
\label{ian-damp}}
\end{figure}

On the other hand, eq. (\ref{B-L}) shows that $k\sim 1$ 
is indeed preferred for explaining the observed magnitude 
of the baryon asymmetry. For $k\ll 1$ the result could be too 
small, since $(B-L)_0\propto k^2$ fastly goes to zero. 

One could question, whether the two sectors would not equilibrate 
their temperatures if $k\sim 1$. 
As far as the mirror sector includes the gauge couplings  
which are the same as the standard ones, 
the mirror particles should be
thermalized at a temperature $ T^\prime$.
Once $ k_1 \leq 1 $, $T^\prime$ will remain smaller than the 
parallel temperature $T$ of the ordinary system, and so 
the presence of the out-of-equilibrium hidden sector does
not affect much the Big Bang nucleosynthesis epoch.

Indeed, if the two sectors had different temperatures at 
reheating, then they evolve independently during
the expansion of the Universe and approach the nucleosynthesis era 
with different temperatures. For $ k_1\leq 1 $, the energy 
density transferred to the mirror sector will be crudely
$ \rho^\prime \approx (8 k_1/g_\ast)\rho$ \cite{BB-PRL},
where $ g_\ast(\approx 100) $ is attained to the leptogenesis epoch.
Thus, translating this to the BBN limits, 
this corresponds to a contribution equivalent to an 
effective number of extra light neutrinos 
$\Delta N_\nu \approx k/14$. 

The following remark is in order. 
The mirror matter could be dark matter even if $k \ll 1$, 
when $Y'_{BL} = Y_{BL}$, if one assumes that the  
M-parity is spontaneously broken so that the 
mirror nucleon masses are about 5 times 
larger than the ordinary ones. 
As we discussed in previous section, this situation would emerge 
if $v'/v \sim 100$, and  $\Lambda'/\Lambda \sim 5$.  
Needless to say, that the considered mechanism is insensitive 
to the values of the weak scale of the mirror sector as far as 
the latter remains much smaller than the masses of the 
heavy singlet neutrinos.

\section{Mirror baryons as dark matter}

We have shown that mirror baryons could provide
a significant contribution to the energy density
of the Universe and thus they could constitute a  
relevant component of dark matter.
An immediate question arises:
how the mirror baryon dark matter (MBDM) behaves
and what are the differences from the more familiar dark
matter candidates as the cold dark matter (CDM),
the hot dark matter (HDM) etc.
In this section we briefly address the
possible observational consequences of
such a cosmological scenario.

In the most general context, the present energy
density contains a relativistic (radiation) component
$\Omega_r$, a non-relativistic (matter) component
$\Omega_m$ and the vacuum energy
density $\Omega_\Lambda$ (cosmological term).
According to the inflationary paradigm the Universe
should be almost flat,
$\Omega_0=\Omega_m + \Omega_r + \Omega_\Lambda \approx 1$,
which agrees well with the recent results
on the CMBR anisotropy and large scale power spectrum.

The Hubble parameter is known to be
$H_0 = 100 h$ km s$^{-1}$ Mpc$^{-1}$ with
$h \approx 0.7$, and for redshifts of cosmological
relevance, $1+z = T/T_0 \gg 1$, it becomes
\be{H}
H(z)= H_0 \left[\Omega_{r}(1+z)^4
+ \Omega_{m} (1+z)^3 + \Omega_\La \right]^{1/2}  .
\end{equation}
In the context of our model, the relativistic fraction
is represented by the ordinary and mirror
photons and neutrinos,
$\Omega_rh^2=4.2\times 10^{-5}(1+x^4)$, and the 
contribution of the mirror species is negligible
in view of the BBN constraint $x< 0.6$.
As for the non-relativistic component,
it contains the O-baryon fraction $\Omega_b$ and
the M-baryon fraction $\Omega'_b = \beta\Omega_b$, 
while the other types of dark matter, e.g. the CDM,
could also be present. Therefore, in general, 
$\Omega_m=\Omega_b +\Omega'_b+\Omega_{\rm cdm}$.\footnote{
In the context of supersymmetry,
the CDM component could exist in the form of
the lightest supersymmetric particle (LSP).
It is interesting to remark that the mass fractions
of the ordinary and mirror LSP are related as
$\Omega'_{\rm LSP} \simeq x\Omega_{\rm LSP}$.
The contribution of the mirror neutrinos
scales as $\Omega'_\nu = x^3 \Omega_\nu$ and thus
it is also irrelevant.
}

The important moments for the structure formation
are related to the matter-radiation equality (MRE) epoch
and to the plasma recombination and matter-radiation
decoupling (MRD) epochs.

The MRE occurs at the redshift 
\be {z-eq}
1+z_{\rm eq}= \frac{\Omega_m}{\Omega_r} \approx
 2.4\cdot 10^4 \frac{\omega_{m}}{1+x^4} 
\end{equation}
where we denote $\omega_m = \Omega_{m}h^2$. 
Therefore, for $x\ll 1$ it is not altered by the additional 
relativistic component of the M-sector. 

The radiation decouples from matter after almost all of
electrons and protons recombine into neutral hydrogen
and the free electron number density  sharply diminishes,
so that the photon-electron scattering rate
drops below the Hubble expansion rate. 
In the ordinary Universe the MRD takes place
in the matter domination period, at the temperature
$T_{\rm dec} \simeq 0.26$ eV, which corresponds to the redshift
$1+z_{\rm dec}=T_{\rm dec}/T_0 \simeq 1100$.

\begin{figure}
\centerline{\psfig{file=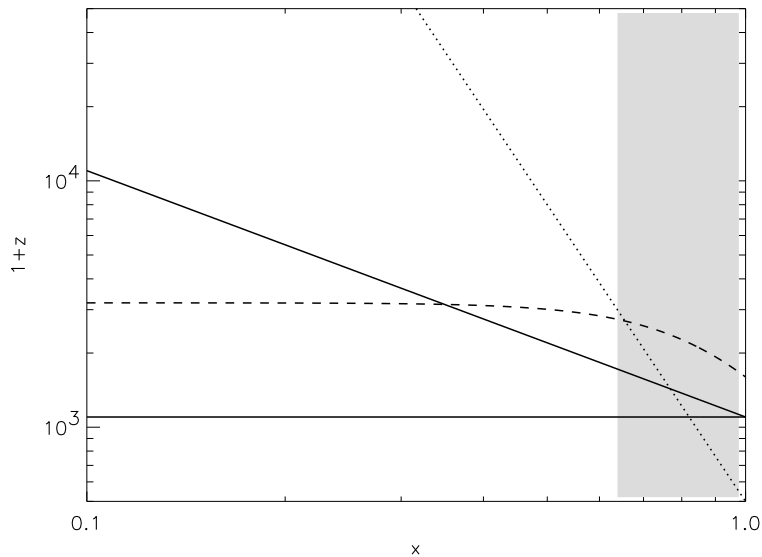,width=10cm}}
\caption{
The M-photon decoupling redshift $1+z_{dec}'$ as a function of $x$ 
(thick solid). The horizontal thin solid line marks the ordinary 
photon decoupling redshift $1+z_{\rm dec} = 1100$. 
We also show the matter-radiation equality redshift
$1+z_{\rm eq}$ (dash) and the mirror Jeans-horizon mass equality
redshift $1+z'_c$ (dash-dot) for the case $\omega_m =0.135$. 
The shaded area $x> 0.64$ is excluded by the BBN limits. 
}
\label{fig3} 
\end{figure}

The MRD temperature in the M-sector $T'_{\rm dec}$
can be calculated following the same lines as in
the ordinary one \cite{BCV}.
Due to the fact that in either case the
photon decoupling occurs when the exponential factor
in Saha equations becomes very small,
we have $T'_{\rm dec} \simeq T_{\rm dec}$,
up to small logarithmic corrections related to
$B'$ different from $B$. Hence
\be{z'_dec}
1+z'_{\rm dec} \simeq x^{-1} (1+z_{\rm dec})
\simeq 1100\, x^{-1}
\end{equation}
so that the MRD in the M-sector occurs earlier
than in the ordinary one. Moreover, for $x$ less than
$x_{\rm eq}=0.045\omega_m^{-1}\simeq 0.3$,
the mirror photons would decouple
yet during the radiation dominated period
(see Fig. \ref{fig3}).

Let us now discuss the cosmological evolution of the MBDM.
The relevant length scale for the gravitational
instabilities is characterized by the mirror Jeans scale
$\lambda'_J \simeq v'_s (\pi/G\rho)^{1/2}$,
where $\rho(z)$ is the matter density at a given redshift
$z$ and $v'_s(z)$ is the sound speed in the M-plasma.
The latter contains more baryons and less photons than the
ordinary one, $\rho'_b=\beta\rho_b$ and
$\rho'_\gamma = x^4\rho_\gamma$.
Let us consider for simplicity the case
when dark matter of the Universe is entirely due to
M-baryons, $\Omega_m\simeq\Omega'_b$. Then we have:
\be{sound}
v'_s(z) \simeq \frac{c}{\sqrt3}
\left(1+ \frac{3\rho'_b}{4\rho'_\gamma}\right)^{-1/2} 
\approx \frac{c}{\sqrt3}
\left[ 1 +\frac34\left(1+x^{-4}\right)
\frac{1+z_{\rm eq}}{1+z}\right]^{-1/2} .   
\end{equation}
Hence, for redshifts 
of cosmological relevance, $z\sim z_{\rm eq}$,
we have $v'_s \sim 2x^2 c/3 \ll c/\sqrt{3}$,
quite in contrast with the ordinary world,
where $v_s \approx c/\sqrt{3}$ practically
until the photon decoupling, $z=1100$.

The M-baryon Jeans mass
$M'_J =\frac{\pi}{6} \rho_m \lambda^{\prime3}_J$ reaches the
maximal value at $z=z'_{\rm dec}\simeq 1100/x$,
$M'_J(z'_{dec}) \simeq 2.4 \cdot 10^{16}\times
x^6 [1+(x_{\rm eq}/x)]^{-3/2}
\omega_m^{-2} ~ M_\odot$.   
Notice, however, that $M'_J$ becomes smaller than the
Hubble horizon mass $M_H = \frac{\pi}{6} \rho H^{-3}$
starting from a redshift
$z_c= 3750 x^{-4} \omega_m $, which is
about $z_{\rm eq}$ for $x=0.64$, but
it sharply increases for smaller values of $x$
(see Fig. \ref{fig3}).
So, the density perturbation scales which enter
the horizon at $z \sim z_{\rm eq}$ have mass larger
than $M'_J$ and thus undergo uninterrupted linear growth
immediately after $t=t_{\rm eq}$.
The smaller scales for which $M'_J > M_H$
instead would first oscillate.
Therefore, the large scale structure
formation is not delayed even if the mirror MRD epoch
did not occur yet, i.e. even if $x> x_{\rm eq}$.
The density fluctuations start to grow in the M-matter
and the visible baryons are involved later, when after 
being recombined they fall into the potential whells 
of developed mirror structures.

Another important feature of the MBDM scenario is that the
M-baryon density fluctuations should undergo  
strong collisional damping around the time of  
M-recombination.
The photon diffusion from the overdense to underdense
regions induce a dragging of charged particles
and wash out the perturbations at scales smaller than the
mirror Silk scale $\lambda'_S \simeq
3\times f(x)\omega_m^{-3/4}$ Mpc,
where $f(x)=x^{5/4}$ for $x > x_{\rm eq}$,
and $f(x) = (x/x_{\rm eq})^{3/2} x_{\rm eq}^{5/4}$
for $x < x_{\rm eq}$.

Thus, the density perturbation scales which can undergo 
the linear growth after the MRE epoch are limited by the
length $\lambda'_S$.
This could help in avoiding the excess of small scales
(of few Mpc) in the power spectrum without
tilting the spectral index.
The smallest perturbations that survive the
Silk damping will have the mass
$M'_S \sim f^3(x) \omega_m^{-5/4} \times 10^{12}~ M_\odot $. 
Interestingly, for $x\sim x_{\rm eq}$ we have
$M'_S \sim 10^{11}~M_\odot$, a typical galaxy mass.   
To some extend, the cutoff effect is
analogous to the free streaming damping in the case of
warm dark matter (WDM), but there are important   
differences. The point is that like usual baryons,
the MBDM should show acoustic oscillations 
whith an impact on the large scale power spectrum.


\begin{figure}[h]
  \begin{center}
    \leavevmode
    \epsfxsize = 11cm
    \epsffile{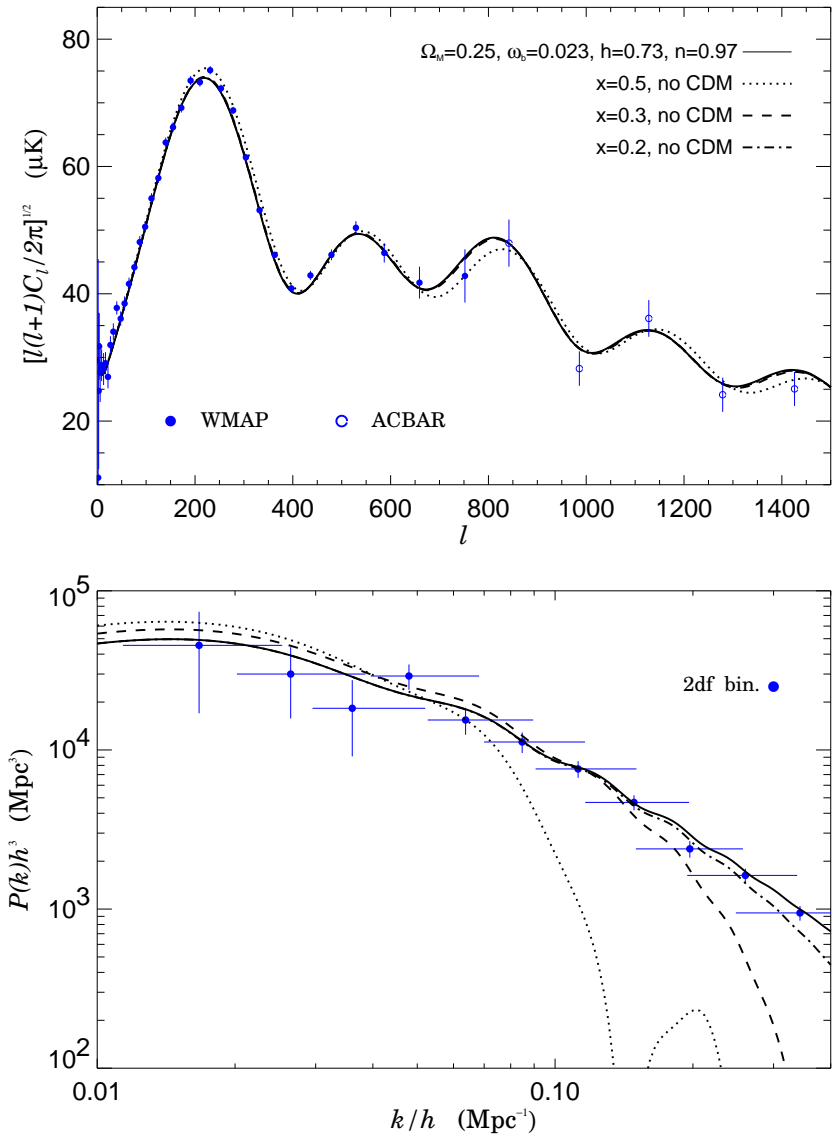}
  \end{center}
\caption{\small The CMBR power spectrum (upper panel) 
and the large scale power spectrum (lower panel) for a 
"concordance" set of cosmological parameters. 
The solid curves correspond to the flat $\Lambda$CDM model, 
while dot, dash and dash-dot curves correspond to 
the situation when the CDM component is completely 
substituted by the MBDM for different values of $x$. 
}
\label{fig4}
\end{figure}

In addition, the MBDM oscillations transmitted
via gravity to the ordinary baryons,
could cause observable anomalies in the CMB
angular power spectrum for $l$'s larger than 200.  
This effect can be observed only if the M-baryon Jeans
scale $\lambda'_J$ is larger than the Silk scale  
of ordinary baryons, 
which sets a principal cutoff for CMB oscillations
around $l\sim 1200$.
As we have seen above, this would require enough large
values of $x$, near the BBN upper bound $x \simeq 0.6$ or so.

If the dark matter is entirely built up by 
mirror baryons, large values of $x$ are 
excluded by the observational data. For the sake 
of demostration, on Fig. \ref{fig4} we show the 
CMBR and LSS power spectra for different values of $x$. 
We see that for $x > 0.3$ the matter power spectrum 
shows a strong deviation from the experimental data. 
This is due to Silk damping effects which suppress 
the small scale power too early, already for 
$k/h\sim 0.2$. However, the values $x<0.3$ are  
compatible with the observational data.

This has a simple explanation. 
Clearly, for small $x$ the M-matter recombines
before the MRE moment, and thus it should rather manifest
as the CDM as far as the large scale structure is concerned.
However, there still can be a crucial difference at
smaller scales which already went non-linear, like galaxies. 
Then one can question whether the MBDM distribution
in halos can be different from that of the CDM?
Namely, simulations show that the CDM forms triaxial
halos with a density profile too clumped towards the
center, and overproduce the small substructures within
the halo. As for the MBDM, it constitutes a sort of
collisional dark matter and thus potentially could avoide
these problems, at least clearly the one related with
the excess of small substructures.

As far as the MBDM constitutes a dissipative dark matter 
like the usual baryons, one would question how it 
can provide extended halos instead of being clumped 
into the galaxy as usual baryons do. 
However, one has to take into account the possibility
that during the galaxy evolution
the bulk of the M-baryons could fastly fragment
into the stars.
A difficult question to address here
is related to the star formation in the M-sector,
also taking into account that its temperature/density 
conditions and chemical contents
are much different from the ordinary ones.
In any case, the fast star formation would
extinct the mirror gas and thus
could avoide the M-baryons to form disk galaxies. 
The M-protogalaxy, which at a certain moment before disk formation
essentially becomes a collisionless system of the  
mirror stars, could maintain a typical elliptical structure.
In other words, we speculate on the possibility
that the M-baryons form mainly elliptical 
galaxies.\footnote{For a comparison, in the ordinary world
the number of spiral and elliptic galaxies are 
comparable. Remarkably, the latter contain old stars, 
very little dust and show low activity of star formation.  
}
Certainly, in this consideration also the galaxy merging
process should be taken into account.
As for the O-matter, within the dark M-matter halo it should
typically show up as an observable elliptic or spiral galaxy,
but some anomalous cases can also be possible,
like certain types of irregular galaxies or even
dark galaxies dominantly made of M-baryons.

\begin{figure}[h]
  \begin{center}
    \leavevmode
    \epsfxsize = 11cm
    \epsffile{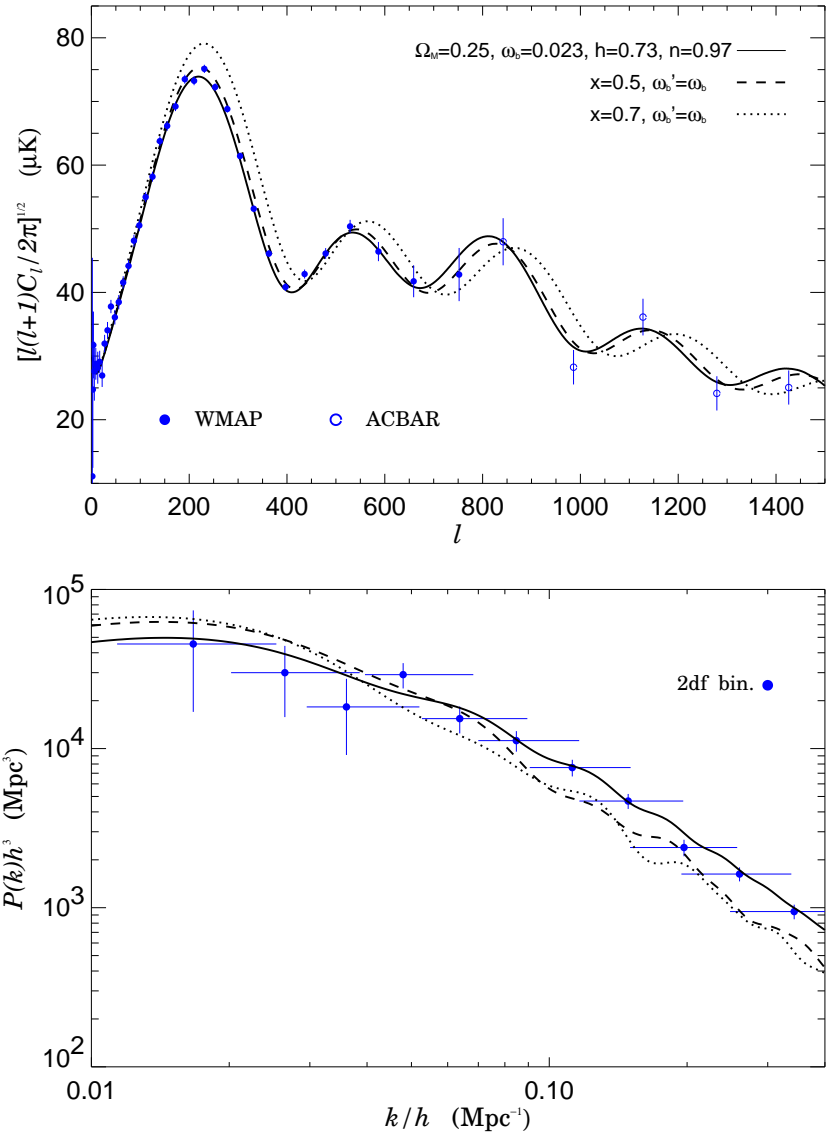}
  \end{center}
\caption{\small The same as on Fig. 4, however for the mixed 
CDM+MBDM scenario for large values of $x$. The ordinary and 
mirror baryon densities are taken equal, $\Om_b'=\Om_B$, 
and the rest of matter density is attained to the SDM component.}
\label{fig5}
\end{figure}

Another tempting issue is whether the M-matter itself
could help in producing big central black holes,
with masses up to $\sim 10^9~ M_\odot$, which are thought
to be the main engines of active galactic nuclei.  

Another possibility can also be considered when 
dark matter in galaxies and clusters contain mixed CDM and
MBDM components, $\Om_d=\Om'_b+\Om_{cdm}$. 
e.g.  one can exploit the case when mirror baryons 
constitute the same fraction of matter as the ordinary ones, 
$\Om'_b=\Om_b$, a situation which emerges naturally in the  
leptogenesis mechanism of sect. 4.3 in the case of small $k$. 

In this case the most interesting and falsificable predictions 
are related to the large $x$ regime. On Fig. \ref{fig5} we show 
the results for the CMBR and LSS power spectra. We see that 
too large values of $x$ are excluded by the CMBR anisotropies, 
but e.g. $x \leq 0.5$ can still be compatible with the data. 

The detailed analysis of this effect will be given elsewhere 
\cite{BCCV}. 
In our opinion, in case of large $x$ the effects on
the CMBR and LSS  can provide direct tests for the MBDM 
and can be falsified by    
the next observations with higher sensitivity.

In the galactic halo
(provided that it is an elliptical mirror galaxy)
the mirror stars should be observed as
Machos in gravitational microlensing  \cite{BDM,Macho}.
Leaving aside the difficult question of the initial
stellar mass function, one can remark that once 
the mirror stars could be very old
and evolve faster than the ordinary ones,
it is suggestive to think that most   
of the massive ones, with mass above the
Chandrasekhar limit $M_{\rm Ch} \simeq 1.5 ~ M_\odot$, 
have already ended up as supernovae, so that only the  
lighter ones 
remain as the microlensing objects.
The recent data indicate the average mass of
Machos around $M\simeq 0.5 ~M_\odot$, which is difficult
to explain in terms of the brown dwarves with masses 
below the hydrogen ignition limit $M < 0.1 M_{\odot}$  
or other baryonic objects \cite{Freese}.
Perhaps, this is the observational evidence
of mirror matter?

It is also possible that in the galactic halo
some fraction of mirror stars exists in the form  
of compact substructures like globular or open clusters.
In this case, for a significant statistics, one could  
observe interesting time and angular correlations
between the microlensing events.

The explosions of mirror supernovae in our galaxy cannot be directly
seen by an ordinary observer. 
However, it should be observed in terms of gravitational waves.
In addition, if the M- and O-neutrinos are mixed \cite{FV,BM},
it can lead to an observable neutrino signal, and could 
be also accompanied by a weak gamma ray burst \cite{GRB}.

\section{Conclusions and outlook}

We have discussed cosmological implications of the   
parallel mirror world with the same microphysics
as the ordinary one, but having smaller temperature,
$T'< T$, with the limit on $x=T'/T<0.6$  set by  
the BBN constraints.
Therefore, the M-sector contains less relativistic
matter (photons and neutrinos) than the O-sector,
$\Omega'_r \ll \Omega_r$.
On the other hand, in the context of certain 
baryogenesis scenarios, the condition
$T'<T$ yields that the mirror sector should produce a
larger baryon asymmetry than the observable one, 
$B'>B$.
So, in the relativistic expansion epoch the cosmological
energy density is dominated by the ordinary component,
while the mirror one gives a negligible contribution.
However, for the non-relativistic epoch
the complementary situation can occur when
the mirror baryon density is bigger
than the ordinary one, $\Omega'_b > \Omega_b$.
Hence, the MBDM can contribute as dark matter along with 
the CDM or even entirely  constitute it.

Unfortunately, we cannot exchange the information
with the mirror physicists and combine our observations.
(After all, as far as the two worlds have the same microphysics,
life should be possible also in the mirror sector, and not 
only $\olga$ but also $\maxim$ could be a real person.)
However, there can be many possibilities to disentangle
the cosmological scenario of two parallel worlds
with the future high precision data concerning  
the large scale structure, CMB anisotropy,
structure of the galaxy halos, gravitational
microlensing, oscillation of neutrinos or other
neutral particles into their mirror partners, etc.

The concept of two parallel worlds is also a sound 
basis for discussion bigravity theories \cite{bigravity}, 
and their cosmological consequences.  

Let us conclude with two quotes of a renowned 
theorist. In 1986 Glashow found a contradiction  
between the estimates of the GUT scale induced kinetic 
mixing term (\ref{FFpr})
and the positronium limits $\eps\leq 4\times 10^{-7}$ 
and concluded that \cite{Glashow86}:  
{\sl "Since these are in evident conflict, the notion of a mirror 
universe with induced electromagnetic couplings of plausible 
(or otherwise detectable) magnitudes is eliminated. The unity 
of physics is again demonstrated when the old positronium workhorse 
can be recalled to exclude an otherwise tenable hypothesis".} 

The situation got another twist within one year, after   
the value $\eps\sim 10^{-7}$ appeared to be interesting   
for tackling the mismatch problem of the orthopositronium 
lifetime. However, in 1987 Glashow has fixed that this value  
was in conflict with the BBN limit $\eps < 3\times 10^{-8}$ 
and concluded the following \cite{Glashow87}:  
{\sl "We see immediately that this limit on $\epsilon$ excludes 
mirror matter as an explanation of the positronium lifetime \dots
We also note that the expected range for $\epsilon$ 
$(10^{-3}-10^{-8})$ is also clearly excluded. 
This suggests that the mirror universe, if it exists at all, 
couples only gravitationally to our own. 
If the temperature of the mirror universe is much lower than 
our own, then no nucleosynthesis limit can be placed on the 
mirror universe at all. 
Then it is also likely that the mirror universe would have 
a smaller baryon number as well, and hence would be virtually empty. 
This makes a hypothetical mirror universe undetectable at energies 
below the Planck energy. 
Such a mirror universe can have no influence on the Earth and 
therefore would be useless and therefore does not exist".}   

In this paper we objected this statement. 
The mirror Universe, if it exists at all, would be useful 
and can have an influence if not directly on the Earth, 
but on the formation of galaxies ... and moreover, 
the very existence of matter, both of visible and dark components,  
can be a consequence of baryogenesis via entropy exchange 
between the two worlds. 
The fact that the temperature of the mirror Universe is much 
lower than the one in our own, does not imply that it would have 
a smaller baryon number as well and hence would be virtually 
empty, but it is likely rather the opposite, 
mirror matter could have larger baryon number and 
being more matter-rich, 
it can provide a plausible candidate for dark matter in the 
form of mirror baryons. Currently it seems to be the 
only concept which could naturally explain the coincidence 
between the visible and dark matter densities of the 
Universe. In this view, future experiments for direct 
detection of mirror matter are extremely interesting. 

Perhaps mirror world is nothing but a reflection of our 
ignorance in the mirror of L-R equivalence.  
Pehaps Alice was wrong and  
there was no real Looking-Glass World beyond the mirror...
However, it's cosmology remains an interesting 
and non-trivial exercise for our imagination, 
which could help in understanding why our 
Universe looks as it is and how it could look 
under other circumstances...











\end{document}